\documentclass[fleqn,10pt]{wlscirep}
\usepackage[utf8]{inputenc}
\usepackage[T1]{fontenc}
\usepackage{bm}
\usepackage{caption}
\usepackage{subcaption}
\usepackage{colortbl}
\usepackage{soul}
\setstcolor{red}

\usepackage{lineno}
\let\oldequation\equation
\let\oldendequation\endequation
\renewenvironment{equation}
  {\linenomathNonumbers\oldequation}
  {\oldendequation\endlinenomath}

\title{\TITLE} 

\author[a]{Moritz Flaschel}
\author[b]{Siddhant Kumar} 
\author[a,*]{Laura De Lorenzis}

\affil[a]{Department of Mechanical and Process Engineering, ETH Z\"{u}rich, 8092 Z\"{u}rich, Switzerland}
\affil[b]{Department of Materials Science and Engineering, Delft University of Technology, 2628 CD Delft, The Netherlands}

\affil[*]{E-mail: ldelorenzis@ethz.ch}

\keywords{unsupervised learning $|$ constitutive models $|$ elasto-plasticity $|$ sparse regression $|$ inverse problems} 

\newcommand{\CA}{\cellcolor{green!15}}
\newcommand{\CB}{\cellcolor{blue!15}}
\newcommand{\CC}{\cellcolor{orange!15}}
\newcommand{\CD}{\cellcolor{purple!15}}

\newcommand{\RR}[1]{\textcolor{black}{#1}}

\newcommand{\ACRO}{EUCLID}
\newcommand{\TITLE}{Discovering \RR{plasticity models} without stress data}

\renewcommand{\figurename}{Fig.~}
\renewcommand{\tablename}{Table~}
\newcommand{\sectionname}{Section~}

\newcommand{\figuresi}[1]{SI Fig.~#1}
\newcommand{\tablesi}[1]{SI Table~#1}
\newcommand{\sectionsi}[1]{SI Section~#1}

\newcommand{\CONSTRAINTS}{Eq.~\eqref{eq:constraint}}
\newcommand{\CFREE}{Eq.~\eqref{eq:Cfree}}
\newcommand{\CDISP}{Eq.~\eqref{eq:Cdisp}}
\newcommand{\CTOTAL}{Eq.~\eqref{eq:objective}}

\newcommand{\idxLOAD}{t}

\newcommand{\idxREAC}{\beta}

\newcommand{\bfepse}{\boldsymbol{\varepsilon}_{\text{e}}}
\newcommand{\bfepsp}{\boldsymbol{\varepsilon}_{\text{p}}}
\newcommand{\bfeps}{\boldsymbol{\varepsilon}}

\newcommand{\rela}{^{\text{rel}}}
\newcommand{\back}{^{\text{back}}}

\newcommand{\fr}{f_r}
\newcommand{\fa}{f_{\alpha}}

\newcommand{\helpA}{\mathbb{C}_{1}} 
\newcommand{\helpB}{\mathbb{C}_{2}} 
\newcommand{\helpC}{\boldface{C}_{3}} 
\newcommand{\helpD}{\mathbb{C}_{4}} 
\newcommand{\helpE}{\boldface{C}_{5}} 
\newcommand{\helpG}{\boldface{C}_{6}} 
\newcommand{\helpH}{C_{7}} 
\newcommand{\helpI}{C_{8}} 
\newcommand{\helpJ}{\mathbb{C}_{9}} 

\newcommand{\ModelVonMisesTrue}{VM}

\newcommand{\ModelSparseOne}{F1}
\newcommand{\ModelSparseTwo}{F2}
\newcommand{\ModelTrescaTrue}{TR}
\newcommand{\ModelTresca}{TR$^*$}
\newcommand{\ModelSchmidtIshlinskyTrue}{SI}
\newcommand{\ModelSchmidtIshlinsky}{SI$^*$}
\newcommand{\ModelIvlevTrue}{IV}
\newcommand{\ModelIvlev}{IV$^*$}
\newcommand{\ModelMariotteTrue}{MA}
\newcommand{\ModelMariotte}{MA$^*$}
\newcommand{\ModelNonConvex}{NC}

\newcommand{\boldface}[1]{\boldsymbol{#1}}  

\newcommand{\bfe}{\boldface{e}}

\newcommand{\bfh}{\boldface{h}}

\newcommand{\bft}{\boldface{t}}
\newcommand{\bfu}{\boldface{u}}
\newcommand{\bfv}{\boldface{v}}

\newcommand{\bfC}{\mathbb{C}}

\newcommand{\bfF}{\boldface{F}}

\newcommand{\bfH}{\boldface{H}}
\newcommand{\bfI}{\boldface{I}}

\newcommand{\bfX}{\boldface{X}}

\newcommand{\bftheta}{\boldsymbol{\theta}}

\newcommand{\bfsigma}{\boldsymbol{\sigma}}

\newcommand{\calD}{\mathcal{D}}

\newcommand{\calN}{\mathcal{N}}

\newcommand{\calS}{\mathcal{S}}

\newcommand{\Rset}{\mathbb{R}}

\newlength{\boxwidth}
\def\dd{\;\!\mathrm{d}}

\def\btheorem{\begin{theorem}}
\def\etheorem{\end{theorem}}
\def\blemma{\begin{lemma}}
\def\elemma{\end{lemma}}
\def\bproposition{\begin{proposition}}
\def\eproposition{\end{proposition}}
\def\bcorollary{\begin{corollary}}
\def\ecorollary{\end{corollary}}
\def\bdefinition{\begin{definition}}
\def\edefinition{\end{definition}}
\def\bexample{\begin{example}}
\def\eexample{\end{example}}
\def\bremark{\begin{remark}}
\def\eremark{\end{remark}}

\def\fg{\boldsymbol}

\newcommand{\be}{\begin{equation}}
\newcommand{\ee}{\end{equation}}
\newcommand{\beq}{\begin{eqnarray}}
\newcommand{\eeq}{\end{eqnarray}}
\newcommand{\bem}{\begin{multline}}
\newcommand{\eem}{\end{multline}}
\newcommand{\ba}{\begin{align}}
\newcommand{\ea}{\end{align}}
\newcommand{\bfzero}{{\fg0}}

\renewcommand{\figurename}{Fig.~}
\renewcommand{\tablename}{Table~}

\begin{abstract}
We propose a new approach for data-driven automated discovery of material laws, which we call EUCLID (Efficient Unsupervised Constitutive Law Identification and Discovery), and we apply it here to the discovery of \RR{plasticity models, including arbitrarily shaped yield surfaces and isotropic and/or kinematic hardening laws}. The approach is \textit{unsupervised}, i.e., it requires no stress data but only full-field displacement and global force data; it delivers \textit{interpretable} models, i.e., models that are embodied by parsimonious mathematical expressions discovered through sparse regression of a potentially large catalogue of candidate functions; it is \textit{one-shot}, i.e., discovery only needs one experiment. The material model library is constructed by expanding the yield function with a Fourier series\RR{, whereas isotropic and kinematic hardening are introduced by assuming a yield function dependency on internal history variables that evolve with the plastic deformation.} For selecting the most relevant Fourier modes \RR{and identifying the hardening behavior}, EUCLID employs physics knowledge, i.e., the optimization problem that governs the discovery enforces the equilibrium constraints in the bulk and at the loaded boundary of the domain. Sparsity promoting regularization is deployed to generate a set of solutions out of which a solution with low cost and high parsimony is automatically selected. Through virtual experiments, we demonstrate the ability of EUCLID to accurately discover several plastic yield surfaces \RR{and hardening mechanisms} of different complexity.
\end{abstract}

\begin{document}
\flushbottom
\maketitle

\thispagestyle{empty}

\section*{Introduction}
Data-driven and machine-learning-based methods are currently pushing forward the frontiers of material modeling. What started with simple regression on uniaxial tensile data has rapidly expanded into high-dimensional and big-data-based surrogate modeling of basically all types of materials of technical interest, including metals, polymers, composites, and more.  While conventional material modeling was based on the a priori assumption of a constitutive law of which the unknown parameters were identified through best-fitting with experimental (or, within multiscale settings, lower-scale computational) results, current data-driven and machine-learning methods give up the usage of an analytical constitutive law altogether. In doing so, they avoid the modeling errors arising due to, e.g., the largely experience-based modeling assumptions and the choice of experimental (or computational) tests being too restrictive to describe the true physics.


Despite the achieved progress, currently available methods are still problematic due to their data-hungry and black-box nature. The state-of-the-art techniques\cite{ghaboussi_knowledgebased_1991,sussman_model_2009,kirchdoerfer_data-driven_2016,ibanez_data-driven_2017,crespo_wypiwyg_2017,gonzalez_learning_2019,mozaffar_deep_2019,zhang_using_2020,kovachki_multiscale_2021} that either \textit{bypass} (directly use data as look-up tables in a model-free fashion) or \textit{surrogate}  (encode in, e.g., artificial neural networks (ANNs) or Gaussian processes (GPs)) material models are rooted in a \textit{supervised
learning} or \textit{curve-fitting} setting. Hence, they need large amount of data consisting of input-output, i.e., strain-stress pairs. Since experimental stress data are only obtainable in the simplest situations,
e.g., uniaxial tensile or bending tests, the comprehensive observation
of strain-stress relations relying on these tests is nearly impossible. Additionally, stress tensors are  challenging to measure experimentally, while   force measurements  only provide incomplete data in the form of boundary-averaged  projections of stress tensors. Multiscale simulations can generate training data sets with tensorial stress-strain pairs, but their computational
cost is still too expensive to probe the entire high-dimensional stress-strain space. Recognition of this issue is very recent.
Within the constitutive-model-free paradigm, it motivated the development of the data-driven identification
method \cite{leygue_data-based_2018,dalemat_measuring_2019}, which formulates the inverse problem associated to the approach in ~\cite{kirchdoerfer_data-driven_2016}. \RR{Likewise recognizing the issue of limited stress data availability, a mathematical framework is proposed in ~\cite{cameron_full-field_2021} to calculate stress fields from deformation fields under the assumption of the alignment of the principal directions of stress and strain or strain rate.} Within the stream of research on surrogating constitutive models with ANNs, recent attempts to use only displacement and
global force data have been performed \cite{tartakovsky_physicsinformed_2020,huang_learning_2020,haghighat_physics-informed_2021}, but are limited to very simple cases (constitutive models of known form with unknown parameters or unknown constitutive models but for one-dimensional cases). Lastly, the \textit{uninterpretability} of stress-strain relations in both paradigms is a standing challenge: it implies significant difficulties in enforcing or verifying the satisfaction of physics constraints and it hinders the extrapolation power.

From the perspectives of both the labeled data requirement and lack of interpretability, the treatment of path-dependent material behavior, such as plasticity, is even more challenging as the stress state at a material point is not solely defined by its strain state, but is additionally dependent on the history of that material point, which is traditionally described using internal variables. Also in this context, the idea of constitutive-model-free approaches is to bypass the formulation of a path-dependent constitutive law, and hence any assumptions on the material behavior, by solving forward problems that are directly informed by the given data\cite{chinesta_data-driven_2017,ibanez_manifold_2018,eggersmann_model-free_2019,tang_map123-ep_2020,carrara_data-driven_2020,carrara_data-driven_2021}.  
The other stream of approaches describe the path-dependent constitutive behavior based on ANNs \cite{mozaffar_deep_2019,li_machine-learning_2019,vlassis_sobolev_2021,huang_machine_2020,shen_prediction_2020,kumar_inverse-designed_2020,zheng_data-driven_2021}, support vector machines \cite{hartmaier_data-oriented_2020}, symbolic regression \cite{bomarito_development_2021} or use the information gained from the data to correct material models known from traditional theories \cite{ibanez_hybrid_2019}. 
Being \textit{supervised}, all these methods require a tremendous amount of labeled data in form of stress-strain  \textit{paths} (with time adding a new dimensionality to the learning problem) for the training process. The additional challenge for path-dependent material behavior is that these paths are needed for several - theoretically infinite - loading histories and often fail to extrapolate beyond the time duration of training data paths. The internal variables of conventional constitutive modeling are physically immeasurable and prohibit any interpretability in the learnt models.

In this light, we propose \ACRO{} (Efficient Unsupervised Constitutive Law Identification and Discovery), an \textit{unsupervised} discovery framework that bridges the advantages of data-driven  and traditional modeling approaches. \ACRO{} does not require the a priori choice of a material model and thus is flexible to describe a variety of different material behaviors, and it only relies on unlabeled data, i.e., full-field displacements (obtained e.g. via Digital Image Correlation (DIC)) and global reaction forces but no stress data, generated by a \textit{single experiment}.
\ACRO{} was recently successfully demonstrated for hyperelastic material model discovery\cite{flaschel_unsupervised_2021} and is here extended to the significantly more challenging problem of  path-dependent elasto-plasticity.
The idea is to formulate a large  library of \textit{interpretable} candidate material models (also referred to as features) and to automatically discover the most relevant features in the library based on the given unlabeled data using only  physics constraints (as opposed to stress labels).
The inspiration for this approach comes from the dynamics community, where sparse regression from a library of candidate features has been used to discover the nonlinear dynamics of physical systems \cite{brunton_discovering_2016}, albeit in a strictly supervised setting. 

\begin{figure*}[tbp]
\centering
\includegraphics[width=\textwidth]{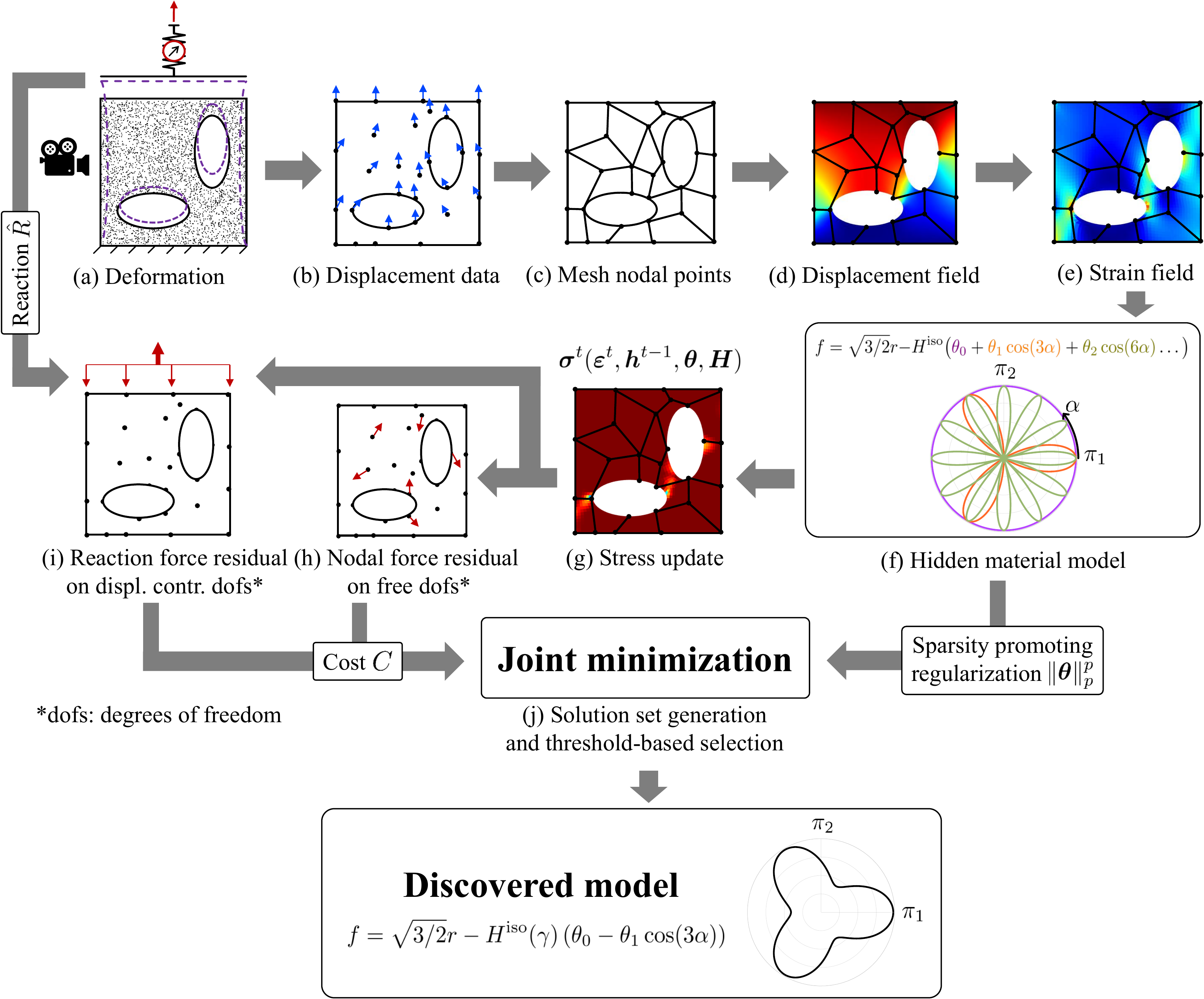}
\caption{Step-by-step schematic of \ACRO. In a single experiment with complex geometry (a), point-wise displacements (b) and global reaction forces (i) are measured. A quadrilateral finite element mesh is constructed (c) to interpolate the displacement data. The resulting displacement field (d) is differentiated to arrive at the strain field (e). The material model library (f) is constructed (here based on a Fourier ansatz). Based on this library and for given material parameters $\bftheta$ \RR{and $\bfH$}, the stresses can be calculated by applying a classical elastic predictor - plastic corrector return mapping algorithm at each load step in the data set, while the history \RR{variables are} updated at each step (g). Based on the stresses, the internal and external virtual works and hence the internal (h) and external (i) force imbalances are calculated, contributing to the cost function $C$. Finally, the cost function is minimized jointly with a sparsity promoting regularization term (j) to generate a set of solutions out of which a solution with low cost and high parsimony is automatically selected. Details are provided in \figuresi{\ref{fig:flowchart}}.
}\label{fig:schematics}
\end{figure*}

Our objective here is to discover fully general \RR{and evolving} plastic yield surfaces\RR{, which characterize the material behavior of three-dimensional elasto-plastic solids,} purely based on \RR{two-dimensional} displacement field and reaction force measurements from only one \RR{experimental} test on an arbitrarily shaped specimen. The material model library is constructed by expanding the yield function with a Fourier series containing a potentially large number of terms. \RR{Yield surface growth and translation, i.e., isotropic and kinematic hardening, are introduced by making the yield function dependent on the accumulated plastic multiplier and the back stress, which are internal history variables that evolve with the plastic deformation.}
For selecting the most relevant Fourier modes \RR{and identifying the hardening mechanisms}, \ACRO{} employs physics knowledge, i.e., the optimization problem that governs the discovery is formulated based on the balance of linear momentum, compensating for the unavailability of stress data.
The step-by-step schematic of \ACRO{} is illustrated in \figurename~\ref{fig:schematics} and described below.

\section*{Results}

\subsection*{Material model library}

In this first work on path-dependent material behavior, we focus on homogeneous, isotropic materials for which linear elastic behavior is followed by \RR{associated, pressure-insensitive plastic behavior with isotropic and/or kinematic hardening}. We also assume small strains and plane stress conditions. In the theory of elasto-plasticity, the infinitesimal strain tensor, which is obtained from the spatial gradient of the displacement field $\bfu$, is additively split into an elastic contribution and a plastic contribution $\bfeps = \bfepse + \bfepsp$, with the plastic strain acting as internal (or history) variable.
The elastic properties of the material are characterized by the stiffness tensor $\bfC$, which determines the linear relation between the elastic strain and the Cauchy stress tensor $\bfsigma = \bfC\colon\bfepse$, whereas the plastic properties are described through the yield function \RR{$f(\bfsigma,\gamma,\bfsigma\back)$, which is here assumed to be dependent on the stress tensor, the accumulated plastic multiplier $\gamma$ and the back stress tensor $\bfsigma\back$}. 
The zero level set \RR{$f=0$} defines the yield surface, i.e., the material deforms elastically if \RR{$f<0$}, and plastic yielding occurs at \RR{$f=0$}. Further, the yield function governs the evolution of the plastic strain through the plastic evolution law
\be\label{eq:evolution}
\dot{\bfeps}_p = \dot{\gamma}\frac{\partial \RR{f(\bfsigma,\gamma,\bfsigma\back)}}{\partial \bfsigma},
\ee
where the superposed dot denotes the derivative with respect to time\RR{. The plastic multiplier and the yield function need} to fulfill the Kuhn-Tucker loading and unloading conditions as well as the consistency condition (see Supplementary Information (SI) Section \ref{sec:model_library_SI}). For details on the theory of elasto-plasticity, the reader is referred to \cite{simo_computational_1998,neto_computational_2008}.

The construction of a suitable material model library, i.e., a large catalogue of potential candidate models, 
builds the basis for the unsupervised discovery framework.
As the elastic material properties can be identified independently from the plastic material properties in a preprocessing step (for example based on the full-field measurements of the first load steps, see \cite{pannier_identification_2006, flaschel_unsupervised_2021}), the elastic stiffness tensor is assumed to be known here and the main focus lies on the discovery of the yield function (see \tablename~\ref{tab:material_models}).
The material model library is constructed by choosing a Fourier series ansatz\footnote{See refs. \cite{ortiz_distortional_1983,raemy_modelling_2017} for other \RR{Fourier-type} expansions of the yield surface.}
\be\label{eq:yield_function}
\RR{f(r , \alpha, \gamma) = \sqrt{\frac{3}{2}} r - H^{\text{iso}}(\gamma) \sum_{i=0}^{n_f} \theta_i \cos(3i\alpha),}
\ee
where $\theta_i$ are the unknown components of the material parameter vector $\bftheta$ and $(n_f+1)$ is the number of features in the library. \RR{The Lode radius $r$ and the Lode angle $\alpha$ are invariants of the relative stress tensor $\bfsigma\rela = \bfsigma - \bfsigma\back$, and are related to the relative principal stresses $\sigma_i$, i.e., the eigenvalues of $\bfsigma\rela$, through}
\be
r = \sqrt{\pi_1^2+\pi_2^2}, \quad \alpha = \text{atan2}(\pi_2,\pi_1),
\quad \text{with}
\quad \pi_1 = {\sqrt{\frac{2}{3}}\sigma_1-\sqrt{\frac{1}{6}}\sigma_2-\sqrt{\frac{1}{6}}\sigma_3,}
\quad \pi_2 = {\sqrt{\frac{1}{2}}\sigma_2-\sqrt{\frac{1}{2}}\sigma_3},
\ee
where $\text{atan2}(\cdot,\cdot)$ is the four-quadrant inverse tangent and the eigenvalues are taken in increasing order, i.e., {$\sigma_1\leq\sigma_2\leq\sigma_3$}. Sine terms as well as certain cosine terms are excluded from the library \eqref{eq:yield_function} to fulfill isotropy requirements (see \sectionsi{\ref{sec:model_library_SI}}).

\RR{Isotropic hardening is considered in \eqref{eq:yield_function} through the nonlinear isotropic hardening function $H^{\text{iso}}(\gamma)$, whereas kinematic hardening is incorporated by letting the back stress evolve nonlinearly with the plastic deformation. We here assume Voce isotropic hardening\cite{voce_relationship_1948,voce_practical_1955} and Armstrong-Frederick kinematic hardening\cite{armstrong_mathematical_1966} by defining}
\RR{
\be
\begin{split}
H^{\text{iso}}(\gamma) &= 1 + H^{\text{iso}}_1 \gamma + H^{\text{iso}}_2 \left(1 - \exp(-H^{\text{iso}}_3 \gamma)\right),\\
\dot{\bfsigma}\back &= H^{\text{kin}}_1 \dot{\bfepsp} - H^{\text{kin}}_2 \dot{\gamma} \bfsigma\back,
\end{split}
\ee
}
\RR{where $\bfH = [ H^\text{iso}_1 \ H^\text{iso}_2 \ H^\text{iso}_3 \ H^\text{kin}_1 \ H^\text{kin}_1 ]$ is a vector containing the unknown hardening parameters that are here assumed to be non-negative.}

By choosing different combinations of active features in the Fourier series,
\eqref{eq:yield_function} can be used to describe smooth yield surfaces of arbitrary shape. 
The representation of the yield function as a closed-form mathematical expression (in contrast to black-box models or constitutive-model-free approaches) facilitates physical interpretation of the material model. E.g., it becomes straightforward to verify whether the yield surface is convex or whether the material behavior is tension-compression symmetric (see \sectionsi{\ref{sec:model_library_SI}}). The closed-form description also enables interpretable constraints on the parameters $\bftheta$ based on physical requirements. \RR{E.g., assuming a vanishing stress state and that no hardening has occurred ($\bfsigma=\boldsymbol{0},\gamma=0,\bfsigma\back=\boldsymbol{0}$), the material is expected to behave elastically ($f<0$), 
\be
\label{eq:constraint}
f(\bfsigma=\boldsymbol{0},\gamma=0,\bfsigma\back=\boldsymbol{0}) = f(r=0,\alpha,\gamma=0) < 0 \implies \theta_0 > \sum_{i=1}^{n_f} |\theta_i|.
\ee
}
The numerical implementation of the model library presented in this section, either for forward finite element simulations (used to generate the artificial data) or for the inverse discovery algorithm (EUCLID), requires the formulation of a stress update procedure. Given the strain $\bfeps^\idxLOAD$ at the current time step $\idxLOAD$, \RR{the history variables $\bfh^{\idxLOAD-1} = \{\bfepsp^{\idxLOAD-1},\gamma^{\idxLOAD-1},(\bfsigma\back)^{\idxLOAD-1}\}$ of the previous time step\footnote{All history variables are assumed to vanish at $t=0$, i.e., $\bfh^0 = \{\bm{0},0,\bm{0}\}$.}
and the material parameters $\bftheta$ and $\bfH$, the current stress  $\bfsigma^\idxLOAD(\bfeps^\idxLOAD,\bfh^{\idxLOAD-1},\bftheta,\bfH)$} is calculated via a classical elastic predictor-plastic corrector return mapping algorithm (see \sectionsi{\ref{sec:model_library_SI}}).

\subsection*{Optimization problem}

To compensate for the unavailability of stress data,
we employ physics knowledge to identify which features in the feature library should be active and to find the values of the corresponding active parameters within $\bftheta$ \RR{and $\bfH$}. 
Under quasi-static loading of a two-dimensional domain $\Omega$ with boundary $\partial\Omega$, the linear momentum balance ($\nabla\cdot\bfsigma=\bfzero$) in its weak formulation is given by
\be
\label{eq:weak}
\int_\Omega \bfsigma^\idxLOAD\RR{(\bfeps^\idxLOAD,\bfh^{\idxLOAD-1},\bftheta,\bfH)}\colon\nabla\bfv \dd A =  \int_{\partial\Omega} \hat \bft^t\cdot \bfv \dd s,
\ee
for all admissible test functions $\bfv$, where $\hat \bft^t$ denotes the boundary tractions.
The available data consists of displacement measurements $\{\bfu^{a,t}:a=1,\dots,n_n; t=1,\dots,n_t\}$ at $n_n$ points and $n_\beta$ net reaction force measurements $\{\hat R^{\beta,t}:\beta=1,\dots,n_\beta; t=1,\dots,n_t\}$ on some  boundary segments, both for $n_t$ time steps. Our objective is to determine $\bftheta$ \RR{and $\bfH$} such that \eqref{eq:weak} is satisfied by the data. \RR{We emphasize that, although a two-dimensional inverse problem is formulated here, the discovered plasticity models are valid for three-dimensional stress states. This means that we exploit a two-dimensional dataset to automatically find plasticity models applicable to three-dimensional solids.}

We create a mesh connecting the points, each point being associated with finite element shape functions $\{N^a(\bfX):a=1,\dots,n_n\}$ such that the strain field is obtained as $\bfeps^t(\bfX)=\sum_a \text{sym}(\nabla N^a (\bfX) \otimes \bfu^{a,t})$, and consequently the stress field $\bfsigma^t(\bfX,\RR{\bfeps^\idxLOAD,\bfh^{\idxLOAD-1},\bftheta,\bfH})$ is determined for each time step via the elasto-plastic constitutive law described by $\bftheta$ \RR{and $\bfH$}. Using the same set of shape functions for the test functions, the  nodal internal forces are computed as
\be
\bfF^{a,t}(\bftheta\RR{,\bfH}) = \int_\Omega \bfsigma^t(\bfX,\RR{\bfeps^\idxLOAD,\bfh^{\idxLOAD-1},\bftheta,\bfH}) \nabla N^a(\bfX) \dd A .
\ee
Equilibrium dictates that the internal forces corresponding to all free degrees of freedom (grouped in the set  $\calD^\text{free}$) be zero at each time step, which naturally leads to the cost function
\be\label{eq:Cfree}
C^\text{free}(\bftheta\RR{,\bfH}) = \sum_{t=1}^{n_t}\sum_{(a,i) \in \calD^\text{free}} \left|F^{a,t}_i(\bftheta\RR{,\bfH})\right|^2.
\ee
At the same time, for each set of constrained degrees of freedom $\calD^{\text{disp},\beta}$, equilibrium requires that the \textit{sum of the internal forces} be balanced at each time step by the corresponding reaction force $\hat{R}^{\beta,t}$,
\be\label{eq:Cdisp}
C^\text{disp}(\bftheta\RR{,\bfH}) = \sum_{t=1}^{n_t} \sum_{\beta=1}^{n_\beta} \left|\hat R^{\beta,t} - \sum_{(a,i)\in \calD^{\text{disp},\beta} }  F^{a,t}_i(\bftheta\RR{,\bfH}) \right|^2.
\ee
We combine the two costs in a single cost function
\be\label{eq:objective}
C(\bftheta\RR{,\bfH}) = C^\text{free}(\bftheta\RR{,\bfH}) + \lambda_r C^\text{\RR{disp}}(\bftheta\RR{,\bfH}).
\ee
with the balancing hyperparameter $\lambda_r>0$. For details, see \sectionsi{\ref{sec:weak_form_SI}}.

The key difference between \ACRO{} and traditional (supervised or unsupervised) parameter identification methods is the fact that the form of the material model is not known \textit{a priori}. In the context of yield surface discovery, the number and combination of active features in \eqref{eq:yield_function} is unknown. Directly minimizing the cost function in \eqref{eq:objective} would 
result in a dense solution vector $\bftheta$, i.e., a highly complex material model with many non-zero material parameters. Such material models are not desired as their calibration and implementation are impractical and they bear a high risk of overfitting the data, possibly resulting in physically inadmissible material behavior. For these reasons, we promote sparsity in the solution vector by formulating an $\ell_{p}$-regularized minimization problem as
\be\label{eq:objective_regularized}
\RR{\{\bftheta^{\text{opt}},\bfH^{\text{opt}}\} = 
\arg\min_{\{\bftheta,\bfH\geq\bf{0}\}}
}
\left(
C(\bftheta\RR{,\bfH})
+ \lambda_p\|\bftheta \|_p^p
\right),
\qquad \text{where} \qquad 
\|\bftheta\|_p = \left(\sum_{i=1}^{n_f}|\theta_i|^p\right)^{1/p}.
\ee
\RR{For the hardening models we assume a mathematical form that is parsimonious at the outset but flexible enough to describe a large class of different hardening mechanisms. To achieve parsimonity in the yield surface, we apply $\ell_{p}$-regularization to the parameter vector $\bftheta$.}
The regularization with hyperparameters $\lambda_p>0$ and $p\in(0,1]$ is a generalization of the LASSO (least absolute shrinkage and selection operator)  \cite{tibshirani_regression_1996} which is recovered for $p=1$. Smaller values of $p$ and higher values of $\lambda_p$ promote sparsity more aggressively, but on the other side increase the degree of non-convexity of the function to be minimized. Here, as in our previous work\cite{flaschel_unsupervised_2021}, we use $p=1/4$, while the choice of $\lambda_p$ is discussed in \sectionsi{\ref{sec:optimization_details_SI}}. Note that in order to fulfill \eqref{eq:constraint}, the parameter $\theta_0$ corresponding to the constant Fourier mode is purposefully not considered in the $\ell_{p}$ regularization. Further, we highlight that applying the $\ell_{p}$ regularization shrinks the absolute values $|\theta_i|$ for $i\in\{1,\dots,n_f\}$ and hence reinforces the fulfillment of the physical constraint in \eqref{eq:constraint}. 
A similar correlation between sparsity of the material model and fulfillment of physical requirements has been observed by \cite{brunton_discovering_2016,flaschel_unsupervised_2021}, supporting the hypothesis that sparse models are more likely to fulfill physical requirements.

\eqref{eq:objective_regularized} is a nonlinear and non-convex optimization problem. As there exists no closed-form expression for the stress-strain relation $\bfsigma^\idxLOAD\RR{(\bfeps^\idxLOAD,\bfh^{\idxLOAD-1},\bftheta,\bfH)}$, it is not feasible to differentiate the cost function in closed form. For this reason, we use a trust-region reflective Newton solver \cite{coleman_convergence_1994} with gradients computed via a finite difference approximation; however, other optimization techniques may also be feasible to use. To tackle the issue of non-convexity, which leads to multiple local minima, we optimize for multiple randomly chosen initial guesses and operate an automatic threshold-based selection which favors a solution with low cost and high parsimony. All details regarding the optimization procedure are provided in \sectionsi{\ref{sec:optimization_details_SI}}.

\begin{figure}[t]
\centering
\includegraphics[width=6cm]{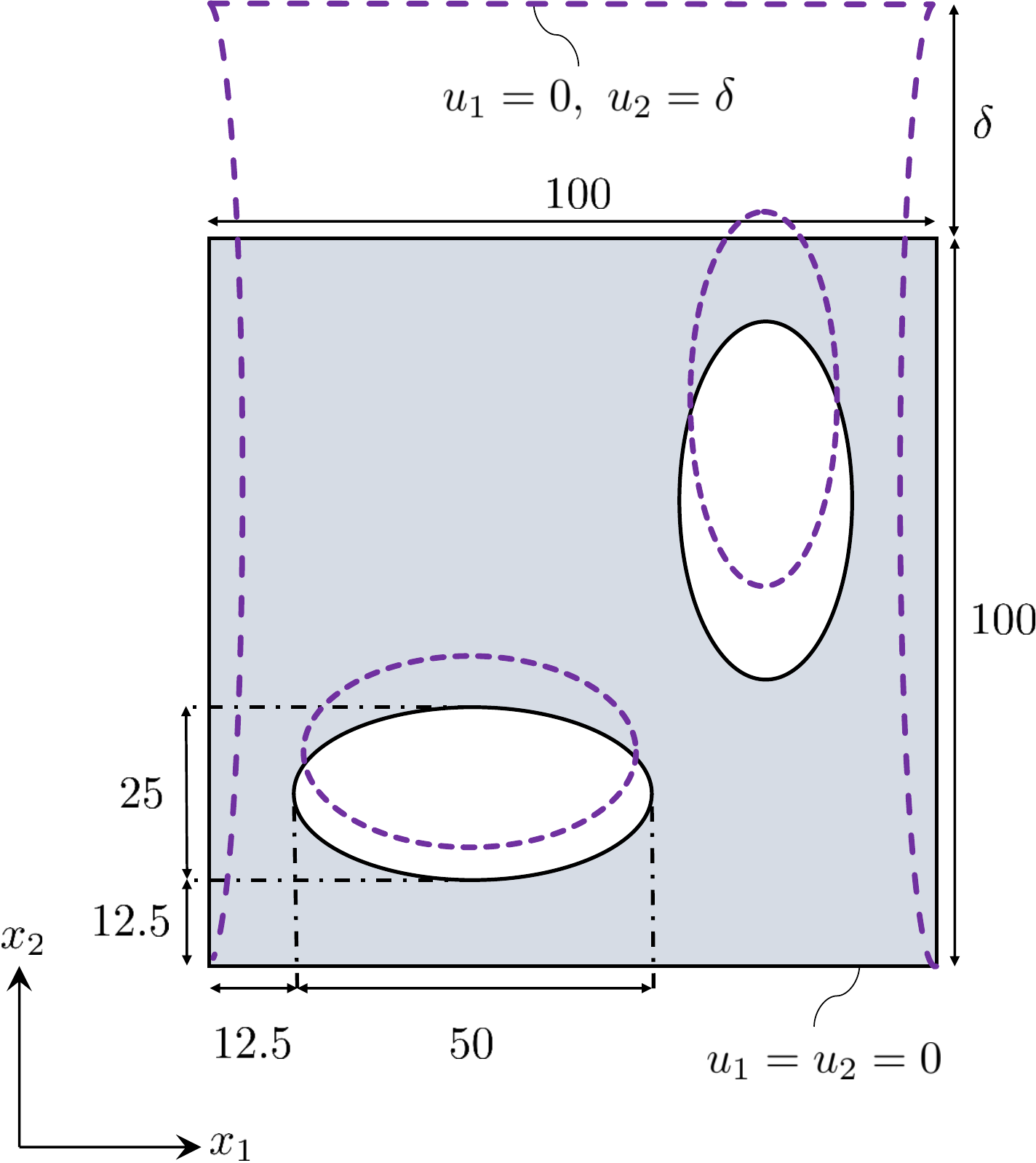}
\caption{Geometry and boundary conditions of the chosen domain: a plate with two elliptic holes under displacement-controlled vertical tension followed by vertical compression. All dimensions are in $\text{mm}$.}
\label{fig:bcs}
\end{figure}

\begin{figure*}[ht]
\centering
\includegraphics[width=14cm]{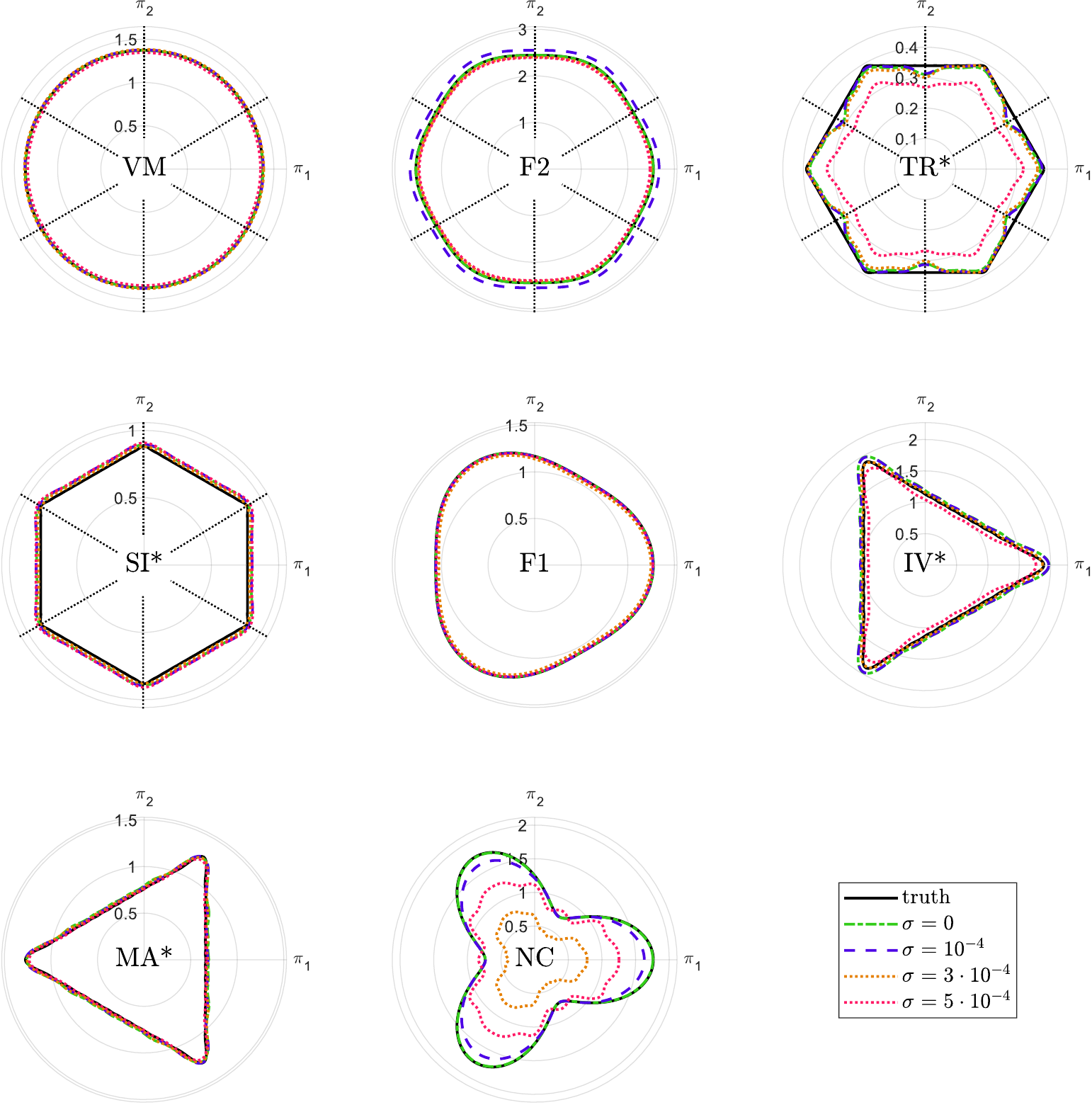}
\caption{Yield surface plots of the (true) hidden and discovered \RR{plasticity} models for different noise levels $\sigma$ (in $\text{mm}$). \RR{Yield surfaces are shown for $\gamma=0.1$, which is an upper limit of the accumulated plastic multiplier in the considered datasets.} Coordinates $\pi_1$, $\pi_2$ are in $\text{kN}/\text{mm}^2$. For the tension-compression symmetric models, the symmetry axes are indicated by black dotted lines.}
\label{fig:yield_surfaces_pi_plane}
\end{figure*}

\begin{figure*}[ht]
\centering
\includegraphics[width=14cm]{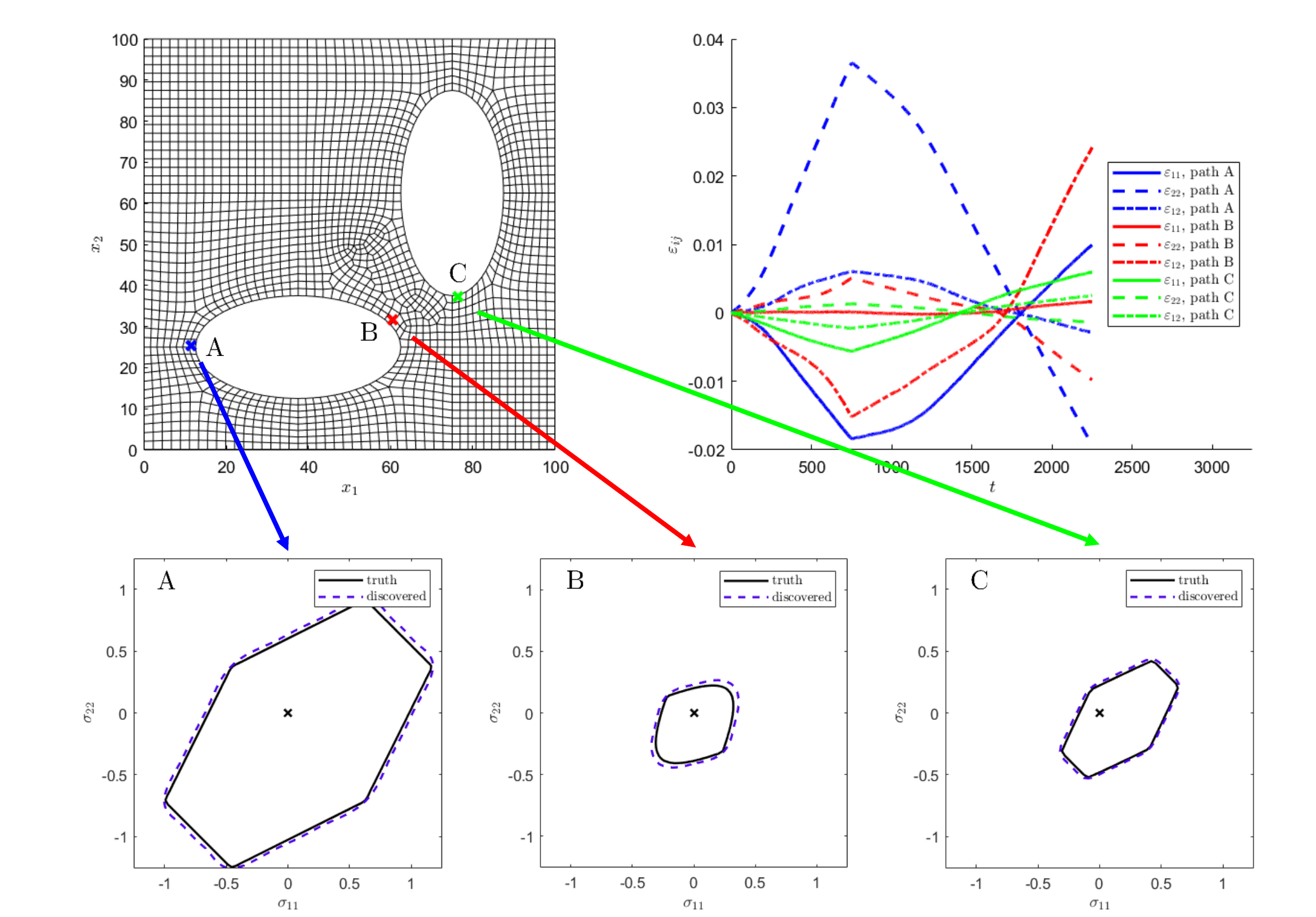}
\caption{
\RR{Yield surface plots of the (true) hidden and discovered Schmidt-Ishlinsky material model (lower row of plots)
at the end of three different deformation paths (A, B, C) corresponding to characteristic points in the specimen domain.
The locations of the points are illustrated in the upper left part of the figure and the strain histories at the different points are shown in the upper right plot.
A noise level of $\sigma=10^{-4}\text{mm}$ was considered and the yield surfaces are plotted at $\sigma_{12}=0$.
Coordinates $x_i$ are in $\text{mm}$ and $\sigma_{ij}$ are in $\text{kN}/\text{mm}^2$.
}}
\label{fig:yield_surfaces_kin_hardening}
\end{figure*}

\begin{table*}[tbp]
\centering
\caption{Yield functions of the (true) hidden and discovered material models for different noise levels $\sigma$ (in $\text{mm}$). \RR{For better clarity, we show the initial yield functions, i.e., $\gamma=0$}. Material parameters $\bftheta$ are in $\text{kN}/\text{mm}^2$. See \tablesi{\ref{tab:material_models_approximations_SI}} for the fully expanded yield function expressions.} 
\label{tab:material_models}
\resizebox{17.6cm}{!}{
\begin{tabular}{lll}
\hline
\multicolumn{2}{c}{Benchmarks} & \multicolumn{1}{l}{Yield Function $f$} \\%
\hline\\[-10pt]
\ModelVonMisesTrue & Truth & $\sqrt{3/2} r - 0.2400$\\%
~ & \CA $\sigma=0$ & \CA $\sqrt{3/2} r - 0.2400$\\%
~ & \CB $\sigma=10^{-4}$ & \CB $\sqrt{3/2} r - 0.2437$\\%
~ & \CC $\sigma=3 \cdot 10^{-4}$ & \CC $\sqrt{3/2} r - 0.2493$\\%
~ & \CD $\sigma=5 \cdot 10^{-4}$ & \CD $\sqrt{3/2} r - 0.2377$\\%
\hline\\[-10pt]
\ModelSparseTwo & Truth & $\sqrt{3/2} r - \left( 0.2350 + 0.0050 \cos(6\alpha) \right)$\\%
~ & \CA $\sigma=0$ & \CA $\sqrt{3/2} r - \left( 0.2350 + 0.0050 \cos(6\alpha) \right)$\\%
~ & \CB $\sigma=10^{-4}$ & \CB $\sqrt{3/2} r - \left( 0.2349 + 0.0055 \cos(6\alpha) \right)$\\%
~ & \CC $\sigma=3 \cdot 10^{-4}$ & \CC $\sqrt{3/2} r - \left( 0.2329 + 0.0047 \cos(6\alpha) \right)$\\%
~ & \CD $\sigma=5 \cdot 10^{-4}$ & \CD $\sqrt{3/2} r - \left( 0.2333 + 0.0041 \cos(6\alpha) \right)$\\%
\hline\\[-10pt]
\ModelTrescaTrue & Truth  & $\max(|\sigma_1-\sigma_2|,|\sigma_2-\sigma_3|,|\sigma_3-\sigma_1|) - 0.2400$\\%
\ModelTresca & Truth$^*$ & $\sqrt{3/2} r - \left( 0.2181 + 0.0127 \cos(6\alpha) + 0.0035 \cos(12\alpha) + 0.0016 \cos(18\alpha) + \dots +  0.0001 \cos(60\alpha) \right)$\\%
~ & \CA $\sigma=0$ & \CA $\sqrt{3/2} r - \left( 0.2129 + 0.0175 \cos(6\alpha) + 0.0033 \cos(18\alpha) \right)$\\%
~ & \CB $\sigma=10^{-4}$ & \CB $\sqrt{3/2} r - \left( 0.2165 + 0.0184 \cos(6\alpha) + 0.0036 \cos(18\alpha) \right)$\\%
~ & \CC $\sigma=3 \cdot 10^{-4}$ & \CC $\sqrt{3/2} r - \left( 0.2290 - 0.0055 \cos(3\alpha) + 0.0208 \cos(6\alpha) - 0.0013 \cos(12\alpha) + 0.0046 \cos(18\alpha) \right)$\\%
~ & \CD $\sigma=5 \cdot 10^{-4}$ & \CD $\sqrt{3/2} r - \left( 0.1388 + 0.0106 \cos(6\alpha) + 0.0017 \cos(12\alpha) + 0.0021 \cos(18\alpha) \right)$\\%
\hline\\[-10pt]
\ModelSchmidtIshlinskyTrue & Truth & $\max(|\sigma_1-(\sigma_2+\sigma_3)/2|,|\sigma_2-(\sigma_3+\sigma_1)/2|,|\sigma_3-(\sigma_1+\sigma_2)/2|) - \cos(\pi / 6) 0.2400$\\%
\ModelSchmidtIshlinsky & Truth$^*$ & $\sqrt{3/2} r - \left( 0.2518 - 0.0146 \cos(6\alpha) + 0.0041 \cos(12\alpha) - 0.0018 \cos(18\alpha) + \dots +  0.0002 \cos(60\alpha) \right)$\\%
~ & \CA $\sigma=0$ & \CA $\sqrt{3/2} r - \left( 0.2260 - 0.0124 \cos(6\alpha) + 0.0028 \cos(12\alpha) \right)$\\%
~ & \CB $\sigma=10^{-4}$ & \CB $\sqrt{3/2} r - \left( 0.2271 - 0.0125 \cos(6\alpha) + 0.0028 \cos(12\alpha) \right)$\\%
~ & \CC $\sigma=3 \cdot 10^{-4}$ & \CC $\sqrt{3/2} r - \left( 0.2374 - 0.0132 \cos(6\alpha) + 0.0030 \cos(12\alpha) \right)$\\%
~ & \CD $\sigma=5 \cdot 10^{-4}$ & \CD $\sqrt{3/2} r - \left( 0.2275 - 0.0130 \cos(6\alpha) + 0.0034 \cos(12\alpha) - 0.0011 \cos(18\alpha) \right)$\\%
\hline\\[-10pt]
\ModelSparseOne & Truth & $\sqrt{3/2} r - \left( 0.2200 + 0.0200 \cos(3\alpha) \right)$\\%
~ & \CA $\sigma=0$ & \CA $\sqrt{3/2} r - \left( 0.2200 + 0.0200 \cos(3\alpha) \right)$\\%
~ & \CB $\sigma=10^{-4}$ & \CB $\sqrt{3/2} r - \left( 0.2225 + 0.0199 \cos(3\alpha) \right)$\\%
~ & \CC $\sigma=3 \cdot 10^{-4}$ & \CC $\sqrt{3/2} r - \left( 0.2177 + 0.0210 \cos(3\alpha) \right)$\\%
~ & \CD $\sigma=5 \cdot 10^{-4}$ & \CD $\sqrt{3/2} r - \left( 0.1915 + 0.0177 \cos(3\alpha) \right)$\\%
\hline\\[-10pt]
\ModelIvlevTrue & Truth & $\max((\sigma_2+\sigma_3)-2\sigma_1,(\sigma_3+\sigma_1)-2\sigma_2,(\sigma_1+\sigma_2)-2\sigma_3) - 0.2400$\\%
\ModelIvlev & Truth$^*$ & $\sqrt{3/2} r - \left( 0.1509 + 0.0415 \cos(3\alpha) + 0.0157 \cos(6\alpha) + 0.0081 \cos(9\alpha) + \dots + 0.0009 \cos(30\alpha) \right)$\\%
~ & \CA $\sigma=0$ & \CA $\sqrt{3/2} r - \left( 0.1458 + 0.0417 \cos(3\alpha) + 0.0157 \cos(6\alpha) + 0.0083 \cos(9\alpha) + 0.0048 \cos(12\alpha) + 0.0032 \cos(15\alpha) + 0.0018 \cos(18\alpha) \right)$\\%
~ & \CB $\sigma=10^{-4}$ & \CB $\sqrt{3/2} r - \left( 0.1490 + 0.0422 \cos(3\alpha) + 0.0165 \cos(6\alpha) + 0.0088 \cos(9\alpha) + 0.0054 \cos(12\alpha) + 0.0029 \cos(15\alpha) + 0.0016 \cos(18\alpha) \right)$\\%
~ & \CC $\sigma=3 \cdot 10^{-4}$ & \CC $\sqrt{3/2} r - \left( 0.1544 + 0.0421 \cos(3\alpha) + 0.0160 \cos(6\alpha) + 0.0084 \cos(9\alpha) + 0.0050 \cos(12\alpha) + 0.0025 \cos(15\alpha) + 0.0015 \cos(18\alpha) \right)$\\%
~ & \CD $\sigma=5 \cdot 10^{-4}$ & \CD $\sqrt{3/2} r - \left( 0.1439 + 0.0427 \cos(3\alpha) + 0.0181 \cos(6\alpha) + 0.0094 \cos(9\alpha) + 0.0039 \cos(12\alpha) + 0.0018 \cos(15\alpha) \right)$\\%
\hline\\[-10pt]
\ModelMariotteTrue & Truth & $\max(\sigma_1-(\sigma_2+\sigma_3)/2,\sigma_2-(\sigma_3+\sigma_1)/2,\sigma_3-(\sigma_1+\sigma_2)/2) - \cos(\pi / 3) 0.2400$\\%
\ModelMariotte & Truth$^*$ & $\sqrt{3/2} r - \left( 0.3018 - 0.0830 \cos(3\alpha) + 0.0315 \cos(6\alpha) - 0.0162 \cos(9\alpha) + \dots + 0.0017 \cos(30\alpha) \right)$\\%
~ & \CA $\sigma=0$ & \CA $\sqrt{3/2} r - \left( 0.1539 - 0.0404 \cos(3\alpha) + 0.0146 \cos(6\alpha) - 0.0071 \cos(9\alpha) + 0.0048 \cos(12\alpha) - 0.0034 \cos(15\alpha) + 0.0019 \cos(18\alpha) \right)$\\%
~ & \CB $\sigma=10^{-4}$ & \CB $\sqrt{3/2} r - \left( 0.1574 - 0.0424 \cos(3\alpha) + 0.0167 \cos(6\alpha) - 0.0083 \cos(9\alpha) + 0.0054 \cos(12\alpha) - 0.0034 \cos(15\alpha) + 0.0017 \cos(18\alpha) \right)$\\%
~ & \CC $\sigma=3 \cdot 10^{-4}$ & \CC $\sqrt{3/2} r - \left( 0.1641 - 0.0445 \cos(3\alpha) + 0.0163 \cos(6\alpha) - 0.0074 \cos(9\alpha) + 0.0043 \cos(12\alpha) - 0.0035 \cos(15\alpha) + 0.0023 \cos(18\alpha) \right)$\\%
~ & \CD $\sigma=5 \cdot 10^{-4}$ & \CD $\sqrt{3/2} r - \left( 0.1664 - 0.0445 \cos(3\alpha) + 0.0177 \cos(6\alpha) - 0.0092 \cos(9\alpha) + 0.0052 \cos(12\alpha) - 0.0029 \cos(15\alpha) + 0.0018 \cos(18\alpha) \right)$\\%
\hline\\[-10pt]
\ModelNonConvex & Truth & $\sqrt{3/2} r - \left( 0.1700 + 0.0700 \cos(3\alpha) \right)$\\%
~ & \CA $\sigma=0$ & \CA $\sqrt{3/2} r - \left( 0.1702 + 0.0699 \cos(3\alpha) \right)$\\%
~ & \CB $\sigma=10^{-4}$ & \CB $\sqrt{3/2} r - \left( 0.1734 + 0.0638 \cos(3\alpha) - 0.0023 \cos(6\alpha) + 0.0013 \cos(9\alpha) \right)$\\%
~ & \CC $\sigma=3 \cdot 10^{-4}$ & \CC $\sqrt{3/2} r - \left( 0.1875 + 0.0615 \cos(3\alpha) - 0.0041 \cos(6\alpha) - 0.0039 \cos(9\alpha) + 0.0052 \cos(12\alpha) + 0.0039 \cos(15\alpha) - 0.0048 \cos(18\alpha) \right)$\\%
~ & \CD $\sigma=5 \cdot 10^{-4}$ & \CD $\sqrt{3/2} r - \left( 0.1958 + 0.0457 \cos(3\alpha) - 0.0038 \cos(6\alpha) - 0.0031 \cos(9\alpha) + 0.0109 \cos(12\alpha) - 0.0022 \cos(15\alpha) - 0.0028 \cos(18\alpha) \right)$\\%
\hline\\[-10pt]
\end{tabular}%
}%
\end{table*}

\begin{table*}[tbp]
\centering
\caption{\RR{Hardening parameters of the (true) hidden and discovered plasticity models for different noise levels $\sigma$ (in $\text{mm}$)}.}
\label{tab:hardening_parameters}
\begin{tabular}{llrrrrr}
\hline
\multicolumn{2}{c}{Benchmarks} & $H^\text{iso}_1$ & $H^\text{iso}_2$ & $H^\text{iso}_3$ & $H^\text{kin}_1$ & $H^\text{kin}_1$ \\%
\hline\\[-10pt]
\ModelVonMisesTrue & Truth & 40.00 & 2.00 & 900.00 & 150.00 & 600.00 \\%
~ & \CA $\sigma=0$ & \CA 40.00 & \CA 2.00 & \CA 900.00 & \CA 150.00 & \CA 600.00 \\%
~ & \CB $\sigma=10^{-4}$ & \CB 39.18 & \CB 2.00 & \CB 846.40 & \CB 148.38 & \CB 626.33 \\%
~ & \CC $\sigma=3 \cdot 10^{-4}$ & \CC 38.01 & \CC 1.99 & \CC 778.13 & \CC 155.19 & \CC 650.66 \\%
~ & \CD $\sigma=5 \cdot 10^{-4}$ & \CD 39.48 & \CD 1.98 & \CD 958.57 & \CD 171.20 & \CD 678.39 \\%
\hline\\[-10pt]
\ModelSparseTwo & Truth & 120.00 & 0.00 & 0.00 & 300.00 & 1000.00 \\%
~ & \CA $\sigma=0$ & \CA 120.00 & \CA 0.00 & \CA 35.46 & \CA 300.00 & \CA 1000.00 \\%
~ & \CB $\sigma=10^{-4}$ & \CB 121.21 & \CB 1.21 & \CB 5.18 & \CB 294.86 & \CB 1014.55 \\%
~ & \CC $\sigma=3 \cdot 10^{-4}$ & \CC 118.65 & \CC 0.02 & \CC 108.71 & \CC 321.29 & \CC 1050.95 \\%
~ & \CD $\sigma=5 \cdot 10^{-4}$ & \CD 117.56 & \CD 0.01 & \CD 12003.28 & \CD 323.39 & \CD 1040.07 \\%
\hline\\[-10pt]
\ModelTresca & Truth & 0.00 & 1.00 & 500.00 & 50.00 & 500.00 \\%
~ & \CA $\sigma=0$ & \CA 0.00 & \CA 0.98 & \CA 525.88 & \CA 56.52 & \CA 545.59 \\%
~ & \CB $\sigma=10^{-4}$ & \CB 0.00 & \CB 0.97 & \CB 498.77 & \CB 52.45 & \CB 563.99 \\%
~ & \CC $\sigma=3 \cdot 10^{-4}$ & \CC 0.00 & \CC 0.82 & \CC 468.30 & \CC 56.45 & \CC 536.35 \\%
~ & \CD $\sigma=5 \cdot 10^{-4}$ & \CD 0.00 & \CD 1.58 & \CD 2859.78 & \CD 74.47 & \CD 579.36 \\%
\hline\\[-10pt]
\ModelSchmidtIshlinsky & Truth & 30.00 & 0.50 & 650.00 & 150.00 & 900.00 \\%
~ & \CA $\sigma=0$ & \CA 30.38 & \CA 0.49 & \CA 570.95 & \CA 140.70 & \CA 813.81 \\%
~ & \CB $\sigma=10^{-4}$ & \CB 30.14 & \CB 0.51 & \CB 519.93 & \CB 139.01 & \CB 833.54 \\%
~ & \CC $\sigma=3 \cdot 10^{-4}$ & \CC 27.39 & \CC 0.50 & \CC 432.26 & \CC 129.30 & \CC 767.45 \\%
~ & \CD $\sigma=5 \cdot 10^{-4}$ & \CD 30.79 & \CD 0.45 & \CD 500.98 & \CD 174.01 & \CD 881.25 \\%
\hline\\[-10pt]
\ModelSparseOne & Truth & 50.00 & 0.50 & 750.00 & 200.00 & 900.00 \\%
~ & \CA $\sigma=0$ & \CA 50.00 & \CA 0.50 & \CA 750.00 & \CA 200.00 & \CA 900.00 \\%
~ & \CB $\sigma=10^{-4}$ & \CB 48.80 & \CB 0.53 & \CB 601.81 & \CB 199.89 & \CB 928.35 \\%
~ & \CC $\sigma=3 \cdot 10^{-4}$ & \CC 48.98 & \CC 0.50 & \CC 782.71 & \CC 210.84 & \CC 911.34 \\%
~ & \CD $\sigma=5 \cdot 10^{-4}$ & \CD 58.27 & \CD 0.60 & \CD 2321.07 & \CD 218.22 & \CD 894.79 \\%
\hline\\[-10pt]
\ModelIvlev & Truth & 75.00 & 1.50 & 1300.00 & 250.00 & 800.00 \\%
~ & \CA $\sigma=0$ & \CA 82.61 & \CA 1.66 & \CA 1192.04 & \CA 257.64 & \CA 824.56 \\%
~ & \CB $\sigma=10^{-4}$ & \CB 78.93 & \CB 1.76 & \CB 1024.17 & \CB 257.66 & \CB 912.86 \\%
~ & \CC $\sigma=3 \cdot 10^{-4}$ & \CC 72.88 & \CC 1.68 & \CC 1101.21 & \CC 249.33 & \CC 865.48 \\%
~ & \CD $\sigma=5 \cdot 10^{-4}$ & \CD 71.87 & \CD 1.66 & \CD 1380.94 & \CD 255.90 & \CD 838.28 \\%
\hline\\[-10pt]
\ModelMariotte & Truth & 40.00 & 1.50 & 800.00 & 200.00 & 850.00 \\%
~ & \CA $\sigma=0$ & \CA 41.79 & \CA 1.64 & \CA 861.47 & \CA 191.87 & \CA 845.45 \\%
~ & \CB $\sigma=10^{-4}$ & \CB 40.26 & \CB 1.58 & \CB 757.54 & \CB 199.24 & \CB 895.49 \\%
~ & \CC $\sigma=3 \cdot 10^{-4}$ & \CC 38.38 & \CC 1.50 & \CC 766.04 & \CC 196.12 & \CC 844.04 \\%
~ & \CD $\sigma=5 \cdot 10^{-4}$ & \CD 36.34 & \CD 1.46 & \CD 742.06 & \CD 208.52 & \CD 833.68 \\%
\hline\\[-10pt]
\ModelNonConvex & Truth & 60.00 & 2.00 & 500.00 & 175.00 & 700.00 \\%
~ & \CA $\sigma=0$ & \CA 59.60 & \CA 2.00 & \CA 498.66 & \CA 174.99 & \CA 694.38 \\%
~ & \CB $\sigma=10^{-4}$ & \CB 55.93 & \CB 1.85 & \CB 475.22 & \CB 234.90 & \CB 890.36 \\%
~ & \CC $\sigma=3 \cdot 10^{-4}$ & \CC 6.44 & \CC 2.24 & \CC 481.16 & \CC 154.28 & \CC 739.98 \\%
~ & \CD $\sigma=5 \cdot 10^{-4}$ & \CD 34.26 & \CD 1.95 & \CD 598.01 & \CD 195.75 & \CD 897.60 \\%
\hline\\[-10pt]
\end{tabular}%
\end{table*}

\subsection*{Benchmarks}
We benchmark EUCLID on \RR{eight} largely different elasto-plastic material models \RR{with different hardening parameters (see \tablename~\ref{tab:material_models},  \tablename~\ref{tab:hardening_parameters} and \figurename~\ref{fig:yield_surfaces_pi_plane} for yield function expressions, hardening parameters and yield surface plots, respectively)}\cite{kolupaev_equivalent_2018,altenbach_extreme_2019}:
\begin{itemize}
\itemsep0em
\item {\ModelVonMisesTrue}: Von-Mises yield function \cite{von_mises_mechanik_1913}
\item {\ModelSparseTwo}: Yield function in \eqref{eq:yield_function} with only $\theta_0,\theta_2 \neq 0$
\item {\ModelTresca}: Smooth approximation of Tresca yield function \cite{tresca_memoire_1869}
\item {\ModelSchmidtIshlinsky}: Smooth approximation of Schmidt-Ishlinsky yield function \cite{schmidt_uber_1932,ishlinsky_hypothesis_1940}
\item {\ModelSparseOne}: Yield function in \eqref{eq:yield_function} with only $\theta_0,\theta_1 \neq 0$
\item {\ModelIvlev}: Smooth approximation of Ivlev yield function \cite{ivlev_theory_1959}
\item {\ModelMariotte}: Smooth approximation of Mariotte yield function \cite{mariotte_traite_1718}
\item {\ModelNonConvex}: Yield function with non-convex yield surface
\end{itemize}
The first \RR{four} material models \ModelVonMisesTrue, \ModelSparseTwo, \ModelTresca, \ModelSchmidtIshlinsky{} exhibit tension-compression symmetry, while the others are tension-compression asymmetric.
Material models \ModelVonMisesTrue, \ModelSparseTwo, \ModelSparseOne, \ModelNonConvex{} are obtained by choosing different combinations of active and inactive parameters in \eqref{eq:yield_function}. \RR{While yield surfaces in classical plasticity are convex and despite the issues connected with the thermodynamic consistency of non-convex yield surfaces, we deliberately choose model \ModelNonConvex{} to evaluate the capabilities of our approach also in the rare case of non-convexity \cite{gluge_does_2018}}.
The original Tresca, Schmidt-Ishlinsky, Ivlev and Mariotte models are characterized by non-smooth yield functions which require complex stress update procedures \cite{neto_computational_2008} beyond the scope of this work. Therefore, and in order to verify the flexibility of the chosen yield function library, the models with non-smooth yield functions are approximated by the Fourier-type expansion in \eqref{eq:yield_function} and denoted by superscript $(\cdot)^*$. Each of the approximated models contains eleven active features with $n_f$ up to 20 in \eqref{eq:yield_function}, see \tablesi{\ref{tab:material_models_approximations_SI}}. The validity of such approximations is proved \RR{in \sectionsi{\ref{sec:validation_Tresca_SI}}.}

\ACRO{} takes full-field displacement and global reaction force measurements as input.
DIC data are emulated by generating artificial data via the finite element method (FEM) based on the material models illustrated above.
The chosen domain is a square plate with two elliptic holes (as schematically shown in \figurename~\ref{fig:bcs}) in plane stress conditions and meshed with bilinear quadrilateral elements.
The plate is deformed under displacement-controlled tension, followed by displacement-controlled compression. In the tension phase, the prescribed vertical displacement $\delta$ is linearly increased from 0 to $\delta = 0.\RR{5}~\text{mm}$, and subsequently decreased to $\delta=-0.\RR{5}~\text{mm}$ in the compression phase.
The nodal displacements are recorded from the FEM solution at a total of $n_t=\RR{2250}$ load steps, $\RR{750}$ load steps in the tension phase and $\RR{1500}$ load steps in the compression phase.
The total horizontal and vertical reaction forces on the top boundary are also recorded.
Note that we purposefully choose a complex specimen geometry, in contrast to the simple geometry of traditional coupon tests, with the objective to obtain a strain field that is rich enough to solve the ill-posed problem of identifying the yield surface with no stress data and just one experiment.

As real DIC data are unavoidably affected by noise in the measured displacement field, we add independent Gaussian noise with zero mean and standard deviation $\sigma>0$ to the synthetic displacement data coming from the FEM simulations (see \sectionsi{\ref{sec:data_generation_SI}}).
We consider noise levels $\sigma \in \{0 \mu\text{m},0.1 \mu \text{m},0.3 \mu \text{m},\RR{0.5 \mu \text{m}}\}$, where $\sigma=0.1~\mu\text{m}$ is considered a reasonable upper limit for modern DIC setups \cite{pierron_extension_2010,marek_extension_2019}. Pushing EUCLID to its breaking point, we further test it for a noise level of $\sigma = 0.5 \mu \text{m}$. The effect of the noise on the yield surface discovery  can be reduced by temporal and spatial smoothing. Here, we restrict ourselves to temporal denoising by applying a Savitzky–Golay filter \cite{savitzky_smoothing_1964} based on quadratic polynomial fitting with a moving-window length of 50 time steps. 

With EUCLID, discovery can proceed from a potentially very large model library - e.g. in our previous work \cite{flaschel_unsupervised_2021} a library with 43 features was used for discovery of hyperelastic strain energy functions. However, it turns out that with the chosen Fourier ansatz a relatively small number of features is already sufficient to provide a remarkably flexible and general yield surface description. Thus, we consider here only \RR{seven} features ($n_f=\RR{6}$) in the model library, i.e., cosine terms with frequencies up to~$\frac{\RR{18}}{2\pi}$. The closed-form expressions of the yield functions discovered by \ACRO{} from the data with different noise levels are reported in \tablename~\ref{tab:material_models}, in comparison to the true expressions. \RR{\tablename~\ref{tab:hardening_parameters} shows the corresponding hardening parameters.} A comparison of the yield surface plots \RR{in the $\pi$ plane} of the true models and the discovered models \RR{after hardening ($\gamma=0.1$)} is presented in \figurename~\ref{fig:yield_surfaces_pi_plane}. \RR{As the $\pi$ plane depends on the relative stresses and not on the absolute stresses, the yield surface plots in \figurename~\ref{fig:yield_surfaces_pi_plane} do not capture kinematic hardening. To this end, yield surface plots in the absolute stress component space are shown in \figurename~\ref{fig:yield_surfaces_kin_hardening} for an exemplary material model at the end of three different deformation paths in the dataset. Similar plots for the other material models are shown in \figuresi{\ref{fig:yield_surfaces_kin_hardening_1234}} and \figuresi{\ref{fig:yield_surfaces_kin_hardening_5678}}.}

In the case of displacements without noise $(\sigma=0)$, material models \ModelVonMisesTrue, \ModelSparseTwo, \ModelSparseOne, \ModelNonConvex{} are discovered exactly, i.e., both the mathematical form of the yield function and the parameters are correctly identified. For increasing noise, the discovered parameters deviate from the true parameters as expected.
\RR{In the non-convex case (\ModelNonConvex{}), false-positive predictions} (features that appear in the discovered formula but are not present in the true model)
are observed.
Material models \ModelTresca, \ModelSchmidtIshlinsky, \ModelIvlev, \ModelMariotte{} on the other hand cannot be described exactly by the chosen model library which makes the exact discovery of the yield function form impossible. However, for the tension-compression symmetric models (\ModelTresca, \ModelSchmidtIshlinsky) it is observed in many cases that tension-compression symmetry-breaking features (e.g., $\cos(3\alpha)$, $\cos(9\alpha)$ \RR{and $\cos(15\alpha)$}) are automatically discarded by \ACRO{}.

Whenever \ACRO{} fails to discover the correct yield function form, parameters ($\theta_i$) corresponding to false-positive features are observed to be considerably smaller than the other parameters and hence have a small influence on the material behavior. Further, false-negative feature predictions \RR{(features that are not discovered although active in the true model)} do not seem to have a high impact on the material behavior either, which is corroborated by the high  accuracy observed in the yield surface plots for all models and noise levels \RR{except higher noise \ModelTresca{} and \ModelNonConvex{}} (see \figurename~\ref{fig:yield_surfaces_pi_plane} \RR{and \figurename~\ref{fig:yield_surfaces_kin_hardening}}). Hence, for the models and noise levels for which \ACRO{} does not find the correct closed-form expression of the yield function, accurate surrogate models are discovered that mimic the behavior of the true yield function. \RR{The Tresca yield surface is the only benchmark for which non-optimal fitting results are observed in shear stress regions even in the case without noise (see \figurename~\ref{fig:yield_surfaces_pi_plane}). The reasons for this are discussed in detail in \sectionsi{\ref{sec:validation_Tresca_SI}}, where we also show that enriching the data with shear deformation drastically improves the results.}

\section*{Discussion}
We show that \ACRO{} is able to discover interpretable \RR{plasticity models} from displacement and net reaction force data only and without using any stress data. The method hence provides a physics-constrained, data-efficient alternative to supervised data-driven and machine learning methods, which require an enormous amount of labelled data and thus are most often inapplicable. The sparse regression enables parsimonious model selection in contrast to an a priori choice of the plasticity model such as in the traditional material model calibration techniques. Hence, after having demonstrated EUCLID for hyperelasticity \cite{flaschel_unsupervised_2021} and for \RR{plasticity, we aim at pursuing its extension to the discovery of more general cases of plasticity with pressure sensitivity and anisotropy} in future work. Further extensions of interest may include other categories of material behavior such as visco-elasticity, visco-plasticity, damage and general combinations thereof. Another important future goal will be the employment of EUCLID on experimental data in the two- and three-dimensional settings using digital image and volume correlation data, respectively.

\section*{Methods}
Detailed method descriptions are provided in the Supplementary Information (SI). \sectionsi{\ref{sec:model_library_SI}} discusses the numerical implementation of the material model library. \sectionsi{\ref{sec:data_generation_SI}} provides details on data generation. The formulation of the objective function and the optimization procedure are discussed in \sectionsi{\ref{sec:weak_form_SI} and \ref{sec:optimization_details_SI}}, respectively.

\section*{Code availability}
The codes generated during the current study are available \RR{at \url{https://euclid-code.github.io/}}.

\section*{Data availability}
The data generated during the current study are available \RR{at \url{https://euclid-code.github.io/}}.

\section*{Author contributions}
M.F., S.K., and L.D.L. conceived the research. M.F.\ developed the theory, implemented the algorithm, and performed the numerical experiments. M.F., S.K., and L.D.L. wrote the paper. 

\section*{Competing interests}
The authors declare no competing interests.

\bibliography{Bib}

\clearpage

\renewcommand\thefigure{S\arabic{figure}}
\renewcommand\thetable{S\arabic{table}}

\setcounter{figure}{0}
\setcounter{table}{0}
\setcounter{equation}{0}

\section*{Supplementary information}


\section{Material model library and numerical implementation aspects} \label{sec:model_library_SI}

\subsection{Candidate material models and physical requirements}
For isotropic material behavior, the yield function can be expressed as a function of the three \RR{relative} principal stresses\footnote{Note that at least one of the \RR{relative} principal stresses is zero for plane stress.} $\sigma_i$ of the \RR{relative} Cauchy stress tensor \RR{$\bfsigma\rela = \bfsigma - \bfsigma\back$ and the accumulated plastic multiplier $\gamma$}
\be
f = f(\sigma_1,\sigma_2,\sigma_3,\RR{\gamma}).
\ee
For pressure-insensitive materials, the yield function does not change along the hydrostatic direction $\sigma_1=\sigma_2=\sigma_3$. It is therefore convenient to apply a coordinate transformation of the \RR{relative} principal stress space 
\be
\begin{pmatrix}
\pi_1\\
\pi_2\\
\pi_3\\
\end{pmatrix}
=
\begin{pmatrix}
\sqrt{\frac{2}{3}} & -\sqrt{\frac{1}{6}} & -\sqrt{\frac{1}{6}}\\
0 & \sqrt{\frac{1}{2}} & -\sqrt{\frac{1}{2}}\\
\sqrt{\frac{1}{3}} & \sqrt{\frac{1}{3}} & \sqrt{\frac{1}{3}}\\
\end{pmatrix}
\begin{pmatrix}
\sigma_1\\
\sigma_2\\
\sigma_3\\
\end{pmatrix},
\ee
such that $\pi_1,\pi_2,\pi_3$ are the transformed stress invariants with $\pi_3$ aligned with the hydrostatic axis.
The yield function can then  be conveniently written as a function of two instead of three \RR{relative} stress invariants \RR{and the accumulated plastic multiplier}
\be
f = f(\pi_1,\pi_2,\RR{\gamma}),
\ee
where the remaining invariants $\pi_1$ and $\pi_2$ span the so-called $\pi$ plane. The yield function can be further rewritten as a function of the polar coordinates $r$ and $\alpha$ in the $\pi$ plane,
\be\label{eq:yield_function_Lode}
f = f(r,\alpha,\RR{\gamma}), \qquad\text{with} \qquad r = \sqrt{\pi_1^2+\pi_2^2},\qquad\alpha = \text{atan2}(\pi_2,\pi_1),
\ee
which are commonly referred to as Lode radius and Lode angle, respectively. Here, $\text{atan2}(\cdot, \cdot)$ denotes the four-quadrant inverse tangent function (also known as two-argument arcus tangent) to correctly identify the polar angle $\alpha$ in the four quadrants of the $\pi$ plane. 

A generalized plastic yield function is chosen by parameterizing \eqref{eq:yield_function_Lode} in terms of a Fourier series 
\be\label{eq:yield_function_Fourier}
f(r,\alpha,\RR{\gamma}) = \sqrt{\frac{3}{2}} r - \RR{H^\text{iso}(\gamma)} \sum_{i=0}^{\infty} \left[ a_i \cos(i\alpha) + b_i \sin(i\alpha) \right],
\ee
where $a_i$ and $b_i$ are real-valued parameters \RR{and $H^\text{iso}(\gamma)$ is a real-valued function (with the property $H^\text{iso}(\gamma=0)=1$) describing the isotropic hardening characteristics}. This parameterization enables the mathematical description of arbitrarily shaped smooth yield surfaces. However,
material isotropy (which we assume here) requires that the yield function is invariant to the order of the \RR{relative} principal stresses, i.e.,
\be
f(\sigma_1,\sigma_2,\sigma_3,\RR{\gamma}) = f(\sigma_2,\sigma_3,\sigma_1,\RR{\gamma}) = f(\sigma_3,\sigma_1,\sigma_2,\RR{\gamma}).
\ee
To fulfill this requirement, parameters $a_i$ in \eqref{eq:yield_function_Fourier} are allowed to be non-zero only for $i\in\{0,3,6,...\}$, while parameters $b_i$ must be zero for all $i$. Defining a new set of real-valued parameters $\theta_i$, we hence arrive at
\be\label{eq:yield_function_SI}
f(r,\alpha,\RR{\gamma}) = \sqrt{\frac{3}{2}} r - \RR{H^\text{iso}(\gamma)} \sum_{i=0}^{n_f} \theta_i \cos(3i\alpha),
\ee
where we truncate the Fourier series after $(n_f+1)$ terms for the numerical treatment. It can further be observed that for tension-compression symmetric material behavior -- which must fulfill $f(r,\alpha=0,\RR{\gamma})=f(r,\alpha=\pi,\RR{\gamma})$ -- only Fourier modes with even-numbered $i$ are allowed.

At $\gamma=0$, the distance $\bar r$ of the yield surface (which is defined through the level set $f=0$) from the origin of the $\pi$ plane ($r=0$) is expressed in terms of the Lode angle $\alpha$ as
\be
\bar r(\alpha) = \RR{\sqrt{\frac{2}{3}}} \sum_{i=0}^{n_f} \theta_i \cos(3i\alpha).
\ee
The yield surface is said to be convex if and only if \cite{gluge_does_2018}
\be
\bar r(\alpha)^2 + 2 \bar r'^2(\alpha) - \bar r(\alpha) \bar r''(\alpha) \geq 0,
\ee
where $\square'$ and $\square''$ denote, respectively, the first and second derivative with respect to $\alpha$.
Considering only two features to be active ($\theta_0 \neq 0$, $\theta_1 \neq 0$), the convexity constraint reduces to $\theta_0 \geq 10 |\theta_1|$ (see \sectionname\ref{sec:optimization_details_SI}).

The yield function governs the evolution of the plastic strain through the plastic evolution law
\be\label{eq:evolution_SI}
\dot{\bfeps}_p = \dot{\gamma}\frac{\partial f}{\partial \bfsigma},
\ee
and has to fulfill the Kuhn-Tucker loading and unloading conditions \cite{simo_computational_1998}
\be\label{eq:Kuhn_Tucker}
f \leq 0, ~ \dot{\gamma} \geq 0, ~ f\dot{\gamma} = 0,
\ee
and the consistency condition
\be
f=0 ~ \Rightarrow ~ \dot{f}\dot{\gamma} = 0.
\ee
\RR{Finally, the isotropic and kinematic hardening models are here chosen by defining $H^{\text{iso}}(\gamma)$ and the back stress evolution law, respectively, as}
\RR{
\be
\begin{split}
H^{\text{iso}}(\gamma) &= 1 + H^{\text{iso}}_1 \gamma + H^{\text{iso}}_2 \left(1 - \exp(-H^{\text{iso}}_3 \gamma)\right),\\
\label{eq:evolution_back}
\dot{\bfsigma}\back &= H^{\text{kin}}_1 \dot{\bfepsp} - H^{\text{kin}}_2 \dot{\gamma} \bfsigma\back,
\end{split}
\ee
}
\RR{where $\bfH = [ H^\text{iso}_1 \ H^\text{iso}_2 \ H^\text{iso}_3 \ H^\text{kin}_1 \ H^\text{kin}_1 ]$ is a vector containing the unknown hardening parameters, which are assumed to be non-negative.}

\subsection{Return mapping algorithm and consistent tangent modulus} \label{sec:return_mapping}
\RR{Collecting all history variables in $\bfh = \{\bfepsp,\gamma,\bfsigma\back\}$,}
we apply an elastic predictor-plastic corrector return mapping algorithm \cite{simo_computational_1998} for the stress update $\bfsigma^\idxLOAD\RR{(\bfeps^\idxLOAD,\bfh^{\idxLOAD-1},\bftheta,\bfH)}$ under plane stress constraints. Assuming plane stress, out-of-plane stress components vanish and out-of-plane strain components can be expressed as functions of the in-plane strain components. In particular, in linear elasticity, 
\be
(\bfepse)_{13}=(\bfepse)_{23}=0,~(\bfepse)_{33} = -\frac{\nu}{1-\nu}\left[(\bfepse)_{11}+(\bfepse)_{22}\right],
\ee
where $\nu$ is Poisson's ratio, and for pressure-insensitive plasticity
\be
(\bfepsp)_{13}=(\bfepsp)_{23}=0,~(\bfepsp)_{33} = -\left[(\bfepsp)_{11}+(\bfepsp)_{22}\right].
\ee
The stress update procedure can hence be implemented for the in-plane components only (see~~\cite{neto_computational_2008}~Section 9.4). First, an elastic predictor trial stress is calculated based on the assumption that the material is fully elastic and there is no evolution of plastic strain, i.e., \RR{the history variables are those known from the previous time step $\bfh^{\idxLOAD,\text{trial}}=\bfh^{\idxLOAD-1}$ (starting from $\bfh^0 = \{\bm{0},0,\bm{0}\}$).} It follows that the trial stress is given by
\be
\bfsigma^{\idxLOAD,\text{trial}} = \bfC (\bfeps^{\idxLOAD}-\bfepsp^{\idxLOAD,\text{trial}}).
\ee
Based on the trial stress, the yield function $f(\bfsigma^{\idxLOAD,\text{trial}}\RR{,\bfh^{\idxLOAD,\text{trial}}})$ is evaluated. Two scenarios are possible. If the yield function is less than or equal to zero, thus fulfilling \eqref{eq:Kuhn_Tucker}, the trial stress is admissible and no plastic correction is needed. However, if the yield function is greater than zero, the trial stress is not admissible and the plastic strain must evolve. The task is then to calculate $\bfepsp^{\idxLOAD}$, $\Delta\gamma^{\idxLOAD}$ \RR{and $(\bfsigma\back)^{\idxLOAD}$} such that $f^\idxLOAD = 0$ and the \RR{evolution laws \eqref{eq:evolution_SI} and \eqref{eq:evolution_back} are} fulfilled. We use an implicit Euler discretization for the \RR{evolution laws} such that
\be\label{eq:evolution_discrete}
\begin{split}
\bfepsp^{\idxLOAD} &= \bfepsp^{\idxLOAD-1} + \Delta\gamma^{\idxLOAD} \frac{\partial f^\idxLOAD}{\partial\bfsigma},\\
(\bfsigma\back)^{\idxLOAD} &= (\bfsigma\back)^{\idxLOAD-1} + H^{\text{kin}}_1 \Delta\gamma^{\idxLOAD} \frac{\partial f^\idxLOAD}{\partial\bfsigma} - H^{\text{kin}}_2 \Delta\gamma^{\idxLOAD} (\bfsigma\back)^{\idxLOAD},
\end{split}
\ee
The nonlinear system of equations above can be linearized and solved iteratively for \RR{$\bfsigma^\idxLOAD$, $\Delta\gamma^{\idxLOAD}$ and $(\Delta\bfsigma\back)^{\idxLOAD}$, followed by the computation of the plastic strain $\bfepsp^{\idxLOAD} = \bfeps^{\idxLOAD} - \bfC^{-1}\colon\bfsigma^\idxLOAD$}.

In addition to calculating $\bfsigma^\idxLOAD$, the implementation of the presented material models in forward finite element simulations (used to generate the data) requires calculating the elasto-plastic consistent tangent modulus
\be
\bfC_{\text{ep}}^{\idxLOAD} = \frac{\partial\bfsigma^\idxLOAD}{\partial\bfeps^\idxLOAD}.
\ee
For $\Delta\gamma^\idxLOAD=0$, it is $\bfC_{\text{ep}}^{\idxLOAD} = \bfC$. \RR{Otherwise, the elasto-plastic consistent tangent modulus for $\Delta\gamma^\idxLOAD>0$ is calculated as (superscripts~$^{\idxLOAD}$ are omitted)
\be
\bfC_{\text{ep}} = \helpJ - \helpI \ (\helpJ \colon \helpE) \otimes \helpG,
\ee
with}
\RR{\begingroup
\allowdisplaybreaks
\begin{align*}
\helpA &= \left( \mathbb{I} - \left( 1 + \Delta\gamma H^{\text{kin}}_2 \right)^{-1} \Delta\gamma H^{\text{kin}}_1 \frac{\partial^2 f}{\partial\bfsigma\partial\bfsigma\back} \right)^{-1} ,\\
\helpB &= \left( 1 + \Delta\gamma H^{\text{kin}}_2 \right)^{-1} \Delta\gamma H^{\text{kin}}_1 \frac{\partial^2 f}{\partial\bfsigma^2} ,\\
\helpC &= - \left( 1 + \Delta\gamma H^{\text{kin}}_2 \right)^{-2} H^{\text{kin}}_2 \left( (\bfsigma\back)^{\idxLOAD-1} + \Delta\gamma H^{\text{kin}}_1 \frac{\partial f}{\partial\bfsigma} \right)
          + \left( 1 + \Delta\gamma H^{\text{kin}}_2 \right)^{-1} \left( H^{\text{kin}}_1 \frac{\partial f}{\partial\bfsigma} + \Delta\gamma H^{\text{kin}}_1 \frac{\partial^2 f}{\partial\bfsigma\partial\gamma} \right) ,\\
\helpD &= \Delta\gamma \frac{\partial^2 f}{\partial\bfsigma^2} + \Delta\gamma \frac{\partial^2 f}{\partial\bfsigma\partial\bfsigma\back} \colon \helpA \colon \helpB ,\\
\helpE &= \frac{\partial f}{\partial\bfsigma} + \Delta\gamma \frac{\partial^2 f}{\partial\bfsigma\partial\gamma} + \Delta\gamma \frac{\partial^2 f}{\partial\bfsigma\partial\bfsigma\back} \colon \helpA \colon \helpC ,\\
\helpG &= \left( \frac{\partial f}{\partial\bfsigma} + \frac{\partial f}{\partial\bfsigma\back} \colon \helpA \colon \helpB \right) \left( \bfC^{-1} + \helpD \right)^{-1} ,\\
\helpH &= \frac{\partial f}{\partial\gamma} + \frac{\partial f}{\partial\bfsigma\back} \colon \helpA \colon \helpC ,\\
\helpI &= \left( \helpG \colon \helpE - \helpH \right)^{-1} ,\\
\helpJ &= \left( \bfC^{-1} + \helpD \right)^{-1} .
\end{align*}
\endgroup}
The tangent modulus is not needed for the inverse problem (\ACRO).

\subsection{Derivatives of the yield function}

As follows, we compute the derivatives of the yield function ($f$) with respect to the stress tensor ($\bfsigma$)\RR{, the accumulated plastic multiplier ($\gamma$) and the back stress ($\bfsigma\back$), which are required for the return mapping algorithm and the consistent tangent modulus.}

The yield function is written as
\be
f = \fr - \RR{H^{\text{iso}}(\gamma)} \fa,
\ee
with
\be
\fr = \sqrt{\frac{3}{2}} r = \sqrt{\frac{3}{2}}\sqrt{\pi_1^2+\pi_2^2}=\sqrt{\frac{3}{2}} \| \bm{\sigma}^{\text{dev}} \|,\qquad \fa=\sum_{i=0}^{n_f} \theta_i \cos(3i\alpha),
\ee
where \RR{$\bfsigma^{\text{dev}} = \bfsigma\rela - 1/3 \ \text{tr}(\bfsigma\rela)\bm{I}$} is the deviatoric part of the \RR{relative} stress tensor and $\bfI$ is the identity tensor.

\subsubsection{Derivatives with respect to the stress tensor}
Noting that
\be
\frac{\partial (\cdot)}{\partial\bfsigma} = \frac{\partial (\cdot)}{\partial\bfsigma\rela},
\ee
we derive in the following the derivative of the yield function with respect to the relative stress.
Employing chain and product rules of differentiation and using the Einstein summation convention, we obtain the first and second derivatives as
\be\label{eq:part_derivs}
\begin{split}
\frac{\partial f}{\partial\sigma_{kl}\rela}
& =  \frac{\partial\fr}{\partial\sigma_{kl}\rela}
- \RR{H^{\text{iso}}(\gamma)} \frac{\partial\fa}{\partial\alpha}
\frac{\partial\alpha}{\partial\sigma_h}
\frac{\partial\sigma_h}{\partial\sigma_{kl}\rela},
\\
\frac{\partial^2 f}{\partial\sigma_{kl}\rela\partial\sigma_{pq}\rela}
& = \frac{\partial^2\fr}{\partial\sigma_{kl}\rela\partial\sigma_{pq}\rela}
- \RR{H^{\text{iso}}(\gamma)} \left( \frac{\partial^2\fa}{\partial\alpha^2}
\frac{\partial\alpha}{\partial\sigma_h}\frac{\partial\sigma_h}{\partial\sigma_{kl}\rela}
\frac{\partial\alpha}{\partial\sigma_g}\frac{\partial\sigma_g}{\partial\sigma_{pq}\rela}
+ \frac{\partial\fa}{\partial\alpha}\frac{\partial^2\alpha}{\partial\sigma_h\partial\sigma_g}\frac{\partial\sigma_h}{\partial\sigma_{kl}\rela}\frac{\partial\sigma_g}{\partial\sigma_{pq}\rela}
+ \frac{\partial\fa}{\partial\alpha}
\frac{\partial\alpha}{\partial\sigma_h}
\frac{\partial^2\sigma_h}{\partial\sigma_{kl}\rela\sigma_{pq}\rela} \right).
\end{split}
\ee
Note that $\sigma_h$ denotes the $h^\text{th}$ \RR{relative} principal stress (with $\sigma_1\leq\sigma_2\leq\sigma_3$), while $\sigma_{kl}\rela$ denotes the $(k,l)$ entry of the \RR{relative} stress tensor $\bfsigma\rela$. The respective partial derivatives in \RR{\eqref{eq:part_derivs}} are provided in the following. The derivatives of $\fr$ with respect to $\bfsigma\rela$ are given by
\be
\begin{split}
\frac{\partial\fr}{\partial\sigma_{kl}\rela}
&= \sqrt{\frac{3}{2}} \frac{\sigma_{kl}^\text{dev}}{\| \bm{\sigma}^{\text{dev}} \|},\\
\frac{\partial^2\fr}{\partial\sigma_{kl}\rela\sigma_{pq}\rela}
&= \sqrt{\frac{3}{2}} \left(\frac{\mathbb{I}^{\text{dev}}_{klpq}}{\| \bm{\sigma}^{\text{dev}} \|}-\frac{{\sigma}^{\text{dev}}_{kl}{\sigma}^{\text{dev}}_{pq}}{\| \bm{\sigma}^{\text{dev}} \|^3}\right),
\end{split}
\ee
where $\mathbb{I}^{\text{dev}}$ is a fourth order tensor defined such that $\bm{\sigma}^{\text{dev}}=\mathbb{I}^{\text{dev}}\colon\bfsigma\RR{\rela}$. The derivatives of $\fa$ with respect to $\alpha$ are given by
\be
\begin{split}
\frac{\partial\fa}{\partial\alpha} &= - \sum_{i=0}^{n_f} 3i \theta_i \sin(3i\alpha),\\
\frac{\partial^2\fa}{\partial\alpha^2} &= - \sum_{i=0}^{n_f} 9i^2 \theta_i \cos(3i\alpha).   
\end{split}
\ee
With $s =1 / \left( (\sigma_1-\sigma_2)^2 + (\sigma_2-\sigma_3)^2 + (\sigma_3-\sigma_1)^2 \right)$, the derivatives of $\alpha$ with respect to the \RR{relative} principal stresses are
\be
\begin{split}
\frac{\partial\alpha}{\partial\sigma_1} &= \sqrt{3}(\sigma_3-\sigma_2)s,\qquad
\frac{\partial^2\alpha}{\partial\sigma_1^2} =
2(-2 \sigma_1 + \sigma_2 + \sigma_3)s\frac{\partial\alpha}{\partial\sigma_1} =
\frac{\partial^2\alpha}{\partial\sigma_2\partial\sigma_3},\\
\frac{\partial\alpha}{\partial\sigma_2} &= \sqrt{3}(\sigma_1-\sigma_3)s,\qquad
\frac{\partial^2\alpha}{\partial\sigma_2^2} =
2(\sigma_1 -2 \sigma_2 + \sigma_3)s\frac{\partial\alpha}{\partial\sigma_2} =
\frac{\partial^2\alpha}{\partial\sigma_1\partial\sigma_3},\\
\frac{\partial\alpha}{\partial\sigma_3} &= \sqrt{3}(\sigma_2-\sigma_1)s,\qquad
\frac{\partial^2\alpha}{\partial\sigma_3^2} =
2(\sigma_1 + \sigma_2 -2 \sigma_3)s\frac{\partial\alpha}{\partial\sigma_3} =
\frac{\partial^2\alpha}{\partial\sigma_1\partial\sigma_2}.
\end{split}
\ee
The derivatives of the \RR{relative} principal stresses with respect to the \RR{relative} stress are computed as
\be
\begin{split}
\frac{\partial\sigma_{i}}{\partial\sigma_{jk}\rela} &= v^{(i)}_j v^{(i)}_k,\\
\frac{\partial^2\sigma_{i}}{\partial\sigma_{jk}\rela\partial\sigma_{lm}\rela} &= v^{(i)}_j \frac{\partial v^{(i)}_k}{\partial\sigma_{lm}\rela} +  \frac{\partial v^{(i)}_j}{\partial\sigma_{lm}\rela} v^{(i)}_k,
\end{split}
\ee
where $\bfv^{(i)}$ denotes the principal direction corresponding to the \RR{relative} principal stress $\sigma_{i}$ and no summation is performed over $i$.
If the \RR{relative} principal stresses are distinct, 
\be
\frac{\partial {v}^{(i)}_j}{\partial\sigma_{lm}\rela} 
= \sum_{\xi\in\{1,2,3\}, \xi \neq i} \frac{
\left({v}^{(\xi)}_p  \frac{\partial \sigma_{pq}\rela}{\partial\sigma_{lm}\rela} {v}^{(i)}_q    \right)
{v}^{(\xi)}_j
}{\sigma_i-\sigma_{\xi}}
= \sum_{\xi\in\{1,2,3\}, \xi \neq i} \frac{
\left({v}^{(\xi)}_l  {v}^{(i)}_m + {v}^{(\xi)}_m  {v}^{(i)}_l \right)
{v}^{(\xi)}_j
}{2(\sigma_i-\sigma_{\xi})},
\ee
For repeated \RR{relative} principal stresses, differentiating the principal directions with respect to the \RR{relative} stress components is not trivial \cite{mills-curran_calculation_1988,wu_improved_2007}. For the sake of simplicity, we approximate summands corresponding to repeated \RR{relative} principal stresses to zero in the equation above, i.e.,
\be
\frac{\partial {v}^{(i)}_j}{\partial\sigma_{lm}\rela} 
= \sum_{\xi\in\{1,2,3\}, \xi \neq i, \sigma_\xi\neq \sigma_i} \frac{
\left({v}^{(\xi)}_l  {v}^{(i)}_m + {v}^{(\xi)}_m  {v}^{(i)}_l \right)
{v}^{(\xi)}_j
}{2(\sigma_i-\sigma_{\xi})}.
\ee
This is a reasonable assumption considering that the chance of two exactly equal \RR{relative} principal stresses is negligible in the numerical simulations.
\RR{\subsubsection{Derivatives with respect to the accumulated plastic multiplier}
We obtain
\be
\frac{\partial f}{\partial\gamma} = - \frac{\partial H^{\text{iso}}(\gamma)}{\partial\gamma} \fa,
\ee
and the mixed derivative
\be
\frac{\partial^2 f}{\partial\bfsigma\partial\gamma} = - \frac{\partial H^{\text{iso}}(\gamma)}{\partial\gamma} \frac{\partial \fa}{\partial\bfsigma},
\ee
where the derivative of $H^{\text{iso}}(\gamma)$ is trivial.
\subsubsection{Derivatives with respect to the back stress}
Employing chain rule and the definition of the back stress, we obtain
\be
\frac{\partial f}{\partial\bfsigma\back} = - \frac{\partial f}{\partial\bfsigma},
\ee
and the mixed derivative}
\be
\RR{\frac{\partial^2 f}{\partial\bfsigma\partial\bfsigma\back} = - \frac{\partial^2 f}{\partial\bfsigma^2}.}
\ee

\subsection{Validation of the Fourier approximation of the Tresca yield surface}\label{sec:validation_Tresca_SI}
The original Tresca, Schmidt-Ishlinsky, Ivlev and Mariotte models are characterized by non-smooth yield functions, which require complex stress update procedures \cite{neto_computational_2008}. To avoid the need for such procedures and in order to verify the flexibility of the chosen yield function library, in this work the models with non-smooth yield functions are approximated by the Fourier-type expansion in \eqref{eq:yield_function_SI} and denoted by superscript $(\cdot)^*$\RR{, see \tablename~\ref{tab:material_models_approximations_SI}}. Each of the approximated models contains eleven active features with $n_f$ up to 20 in \eqref{eq:yield_function_SI}.

\begin{figure}
\begin{center}
\includegraphics[width=0.6\textwidth]{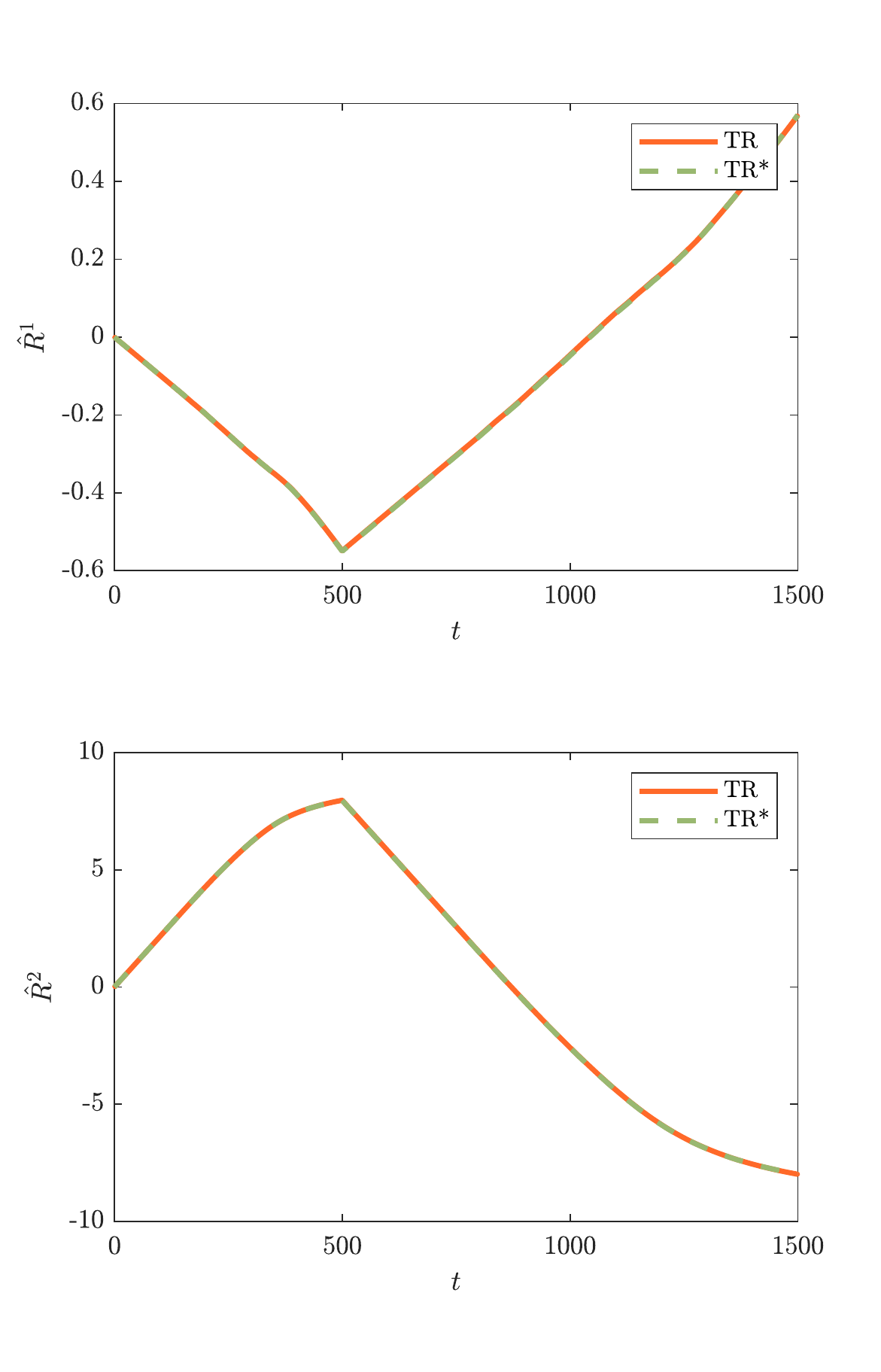}
\caption{Simulated net reaction forces considering the Tresca material model (\ModelTrescaTrue) and its approximation in terms of the Fourier series (\ModelTresca). $\hat{R}^1$ and $\hat{R}^2$ denote the horizontal and vertical reaction force (in $\text{kN}$) at the upper specimen boundary, respectively.}
\label{fig:reaction_force_Tresca}
\end{center}
\end{figure}

\begin{figure}
\centering
\includegraphics[width=\textwidth]{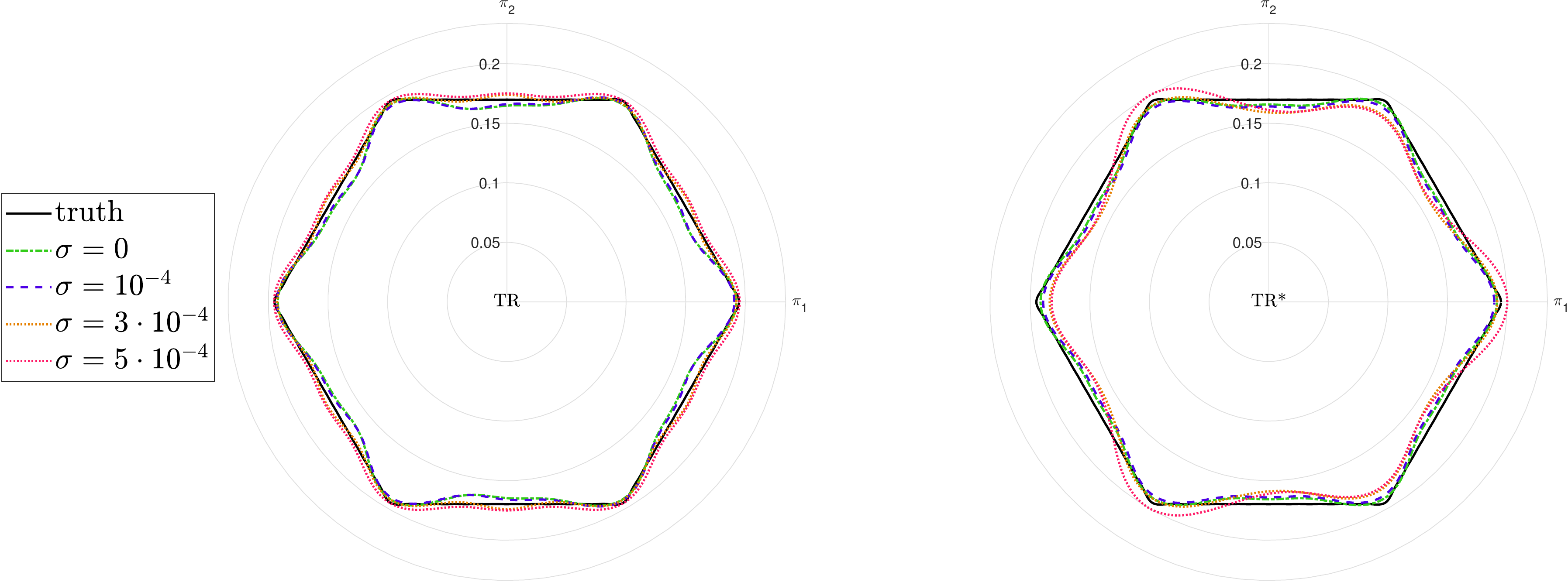}
\caption{\RR{Yield surface plot of the Tresca model (\ModelTrescaTrue{}), its Fourier-type approximation (\ModelTresca{}) and the discovered material models for different noise levels. Coordinates $\pi_1$, $\pi_2$ are in $\text{kN}/\text{mm}^2$.}}
\label{fig:yield_surfaces_Tresca}
\end{figure}

\RR{In this section, we prove the validity of such approximations exemplarily on the Tresca model. As the Fourier-type approximation only introduces errors on the shape of the yield surface but not on its hardening behavior, we show the validation for the perfectly plastic Tresca model ($\bfH = \bm{0}$). First, we show that the Tresca model \ModelTrescaTrue{} and its Fourier-type approximation \ModelTresca{} yield the same predictions in forward finite element computations. To this end, we simulate the deformation of the specimen described in the main article in a tension phase (up to $\delta = 0.1$) and consecutive compression phase (up to $\delta = -0.1$) based on the models \ModelTrescaTrue{} and \ModelTresca{}, and we compute the net reaction forces at the upper boundary of the specimen. As shown in \figurename~\ref{fig:reaction_force_Tresca}, a perfect match between the two is obtained. Second, we apply the discovery algorithm, which is based on the smooth Fourier library, to the data generated from the models \ModelTrescaTrue{} and \ModelTresca{} and show that the discovered models are in good agreement with the true models, see \figurename~\ref{fig:yield_surfaces_Tresca}.}

\begin{figure}
\centering
\includegraphics[width=0.5\textwidth]{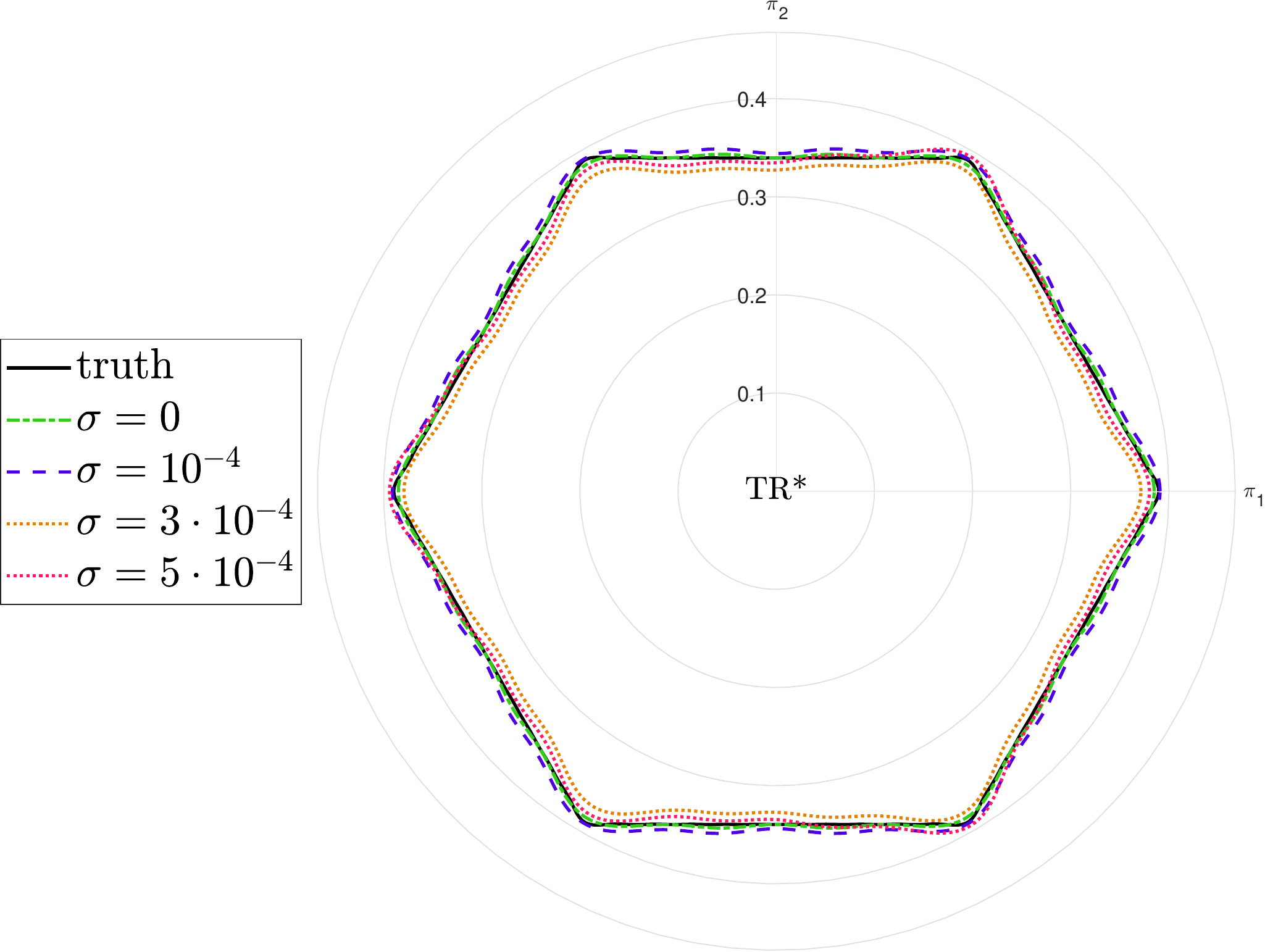}
\caption{\RR{Yield surface plots of the (true) hidden and discovered Tresca model (\ModelTresca{}) based on the shear data enriched dataset for different noise levels. Yield surfaces are shown for $\gamma=0.1$. Coordinates $\pi_1$, $\pi_2$ are in $\text{kN}/\text{mm}^2$.}}
\label{fig:yield_surfaces_Tresca_shear}
\end{figure}

\RR{As can be seen in the main article results as well as in \figurename~\ref{fig:yield_surfaces_Tresca}, non-optimal fitting accuracy is obtained for the Tresca yield surface in shear stress regions even in the case without noise. The reason for this is that the considered (artificial) experimental loading setup generates comparatively low shear deformations in the specimen. As the Tresca material model has a low shear stress resistance, the lack of significant shear deformation data is more detrimental for the Tresca model compared to the other plasticity models. We show in \figurename~\ref{fig:yield_surfaces_Tresca_shear} that enriching the dataset with shear deformation drastically improves the results for the Tresca model \ModelTresca{} with mixed isotropic and kinematic hardening. To this end, a horizontal displacement of the upper specimen boundary was assumed during the data generation in addition to the vertical displacements.}

\section{Data generation}\label{sec:data_generation_SI}
Two-dimensional full-field displacement measurements, e.g., obtained through Digital Image Correlation (DIC), are assumed to be known. In this work, artificial data is generated through simulations with the finite element method (FEM), see \tablename~\ref{tab:settings_SI}. To emulate real experiments that include measurement noise, we add independent normally distributed noise with zero mean and standard deviation $\sigma>0$ to the simulated displacements,
\be
u^{a,\idxLOAD}_i = u^{\text{fem},a,\idxLOAD}_i + u^{\text{noise},a,\idxLOAD}_i, \quad u^{\text{noise},a,\idxLOAD}_i \sim \calN(0,\sigma) \quad \forall \quad a\in\{1,\dots,n_n\},~i\in\{1,2\},~\idxLOAD\in\{1,\dots,n_t\},
\ee
where $u^{\text{fem},a,\idxLOAD}_i$ is the $i^{\text{th}}$ component of the displacement at node $a$ at the $t^\text{th}$ time step obtained from  FEM, and $u^{\text{noise},a,\idxLOAD}_i$ denotes the noise added to it. After applying artificial noise to the data, temporal Savitzky–Golay denoising \cite{savitzky_smoothing_1964} with quadratic polynomial order and a moving-window length of 50 
time steps is applied. 

\section{Weak formulation of linear momentum balance and choice of test functions}\label{sec:weak_form_SI}
To facilitate derivatives and integrals of the field quantities, we associate the points at which the displacements are known to a finite element mesh. For the purpose of this work, we use 4-node bilinear quadrilateral elements with four quadrature points each. The displacement field is then approximated as
\be\label{eq:discretization_u}
\bfu^\idxLOAD(\bfX) = \sum_{a=1}^{n_n}N^a(\bfX) \,\bfu^{a,\idxLOAD},
\ee
where $N^a:\Omega \rightarrow \Rset$ is the shape function associated with the  $a^\text{th}$ node. Under the small strain assumption, the infinitesimal strain tensor is calculated from the displacement field by
\be
\bfeps^\idxLOAD(\bfX)
= \nabla^{\text{sym}}\bfu^\idxLOAD(\bfX)
= \sum_{a=1}^{n_n} \frac{1}{2} \left[\bfu^{a,\idxLOAD} \otimes \nabla N^a(\bfX) + \nabla N^a(\bfX) \otimes \bfu^{a,\idxLOAD}\right],
\ee
where $\nabla$ denotes the gradient operator with respect to the reference coordinates and $\nabla^{\text{sym}}=\frac{1}{2}(\nabla+\nabla^T)$ denotes the symmetric gradient operator.

Let $\calD=\{(a,i) : a=1,\dots,n_n;i=1,2\}$ denote the set of all nodal degrees of freedom where $n_n$ is the number of nodes at which displacements are known. $\calD$ is further split into two non-intersecting subsets of free and displacement-constrained degrees of freedom: $\calD^{\text{free}}$ and $\calD^{\text{disp}}$, respectively, such that $\calD^{\text{free}}\cup\calD^{\text{disp}}=\calD$ and $\calD^{\text{free}}\cap\calD^{\text{disp}}=\emptyset$.

On the basis of the interpolated strain field and the material model library (\sectionname\ref{sec:model_library_SI}), the objective is to find suitable material parameters $\bm{\theta}$ \RR{and $\bfH$} such that linear momentum balance  is fulfilled both in the bulk material and at the boundary. Note that the angular momentum balance is automatically fulfilled in the described setting. Assuming negligible body forces, the weak formulation of linear momentum balance in the reference domain $\Omega$ under quasi-static conditions reads
\be\label{eq:weak_form}
\int_\Omega \bfsigma^\idxLOAD\RR{(\bfeps^\idxLOAD,\bfh^{\idxLOAD-1},\bftheta,\bfH)}\colon\nabla \bfv \dd A - \int_{\partial\Omega}\hat \bft^\idxLOAD \cdot \bfv \dd S = 0, \quad \forall \quad  \text{admissible test function} \ \bfv,
\ee
where $\hat \bft$ is the boundary traction. The test functions $\bfv$ are admissible if they are sufficiently regular and vanish at the fixed degrees of freedom. Note that we prefer the weak to the strong formulation of linear momentum balance because the double spatial derivatives required by the latter make it more sensitive to noise.

Using for $\bfv$ the same approximation basis as for $\bfu$ (Bubnov-Galerkin method), the test functions are approximated as
\be
\bfv(\bfX) = \sum_{a=1}^{n_n}N^a(\bfX) \,\bfv^a,
\ee
and are assumed constant over time. Therefore, \eqref{eq:weak_form} reduces to
\be
\sum_{a=1}^{n_n}\bfv^a\cdot\left[ \underbrace{\int_\Omega \bfsigma^\idxLOAD\RR{(\bfeps^\idxLOAD,\bfh^{\idxLOAD-1},\bftheta,\bfH)}\nabla N^a(\bfX) \dd A}_{\bfF^{a,t}} - \int_{\partial\Omega} {\hat\bft}^t N^a(\bfX) \dd S\right] = 0, \quad \forall \quad  \text{admissible} \ \bfv^a.
\ee
where $\bfF^{a,t}$ is interpreted as the nodal internal force on the $a^\text{th}$ node at time step $t$.
At each interior node as well as at each node located on the unrestrained portion of the boundary, i.e., for each degree of freedom in $\calD^\text{free}$, the second integral vanishes and \eqref{eq:weak_form} yields
\be\label{eq:weak_bulk}
F^{a,t}_i(\bftheta\RR{,\bfH}) = 0,\quad \forall\ (a,i)\in\calD^\text{free}.
\ee
This can also be interpreted as satisfying the weak form of linear momentum balance  for the following virtual fields $\bfv$:
\be
\eqref{eq:weak_form}\quad \text{such that}\quad \bfv = N^a(\bfX){\hat\bfe}^i, \quad   \forall\ (a,i)\in\calD^\text{free},
\ee
where $\{{\hat\bfe}^1, {\hat\bfe}^2\}$ are the cartesian coordinate axes.

In general, the reaction force is not known at every displacement-constrained degree of freedom in $\calD^{\text{disp}}$. Instead, only the sums of the reaction forces for subsets of the displaced degrees of freedom (each subset corresponding to one displaced side of the specimen and one coordinate direction) are known, emulating force measurements in an experiment. Let $n_{\idxREAC}$ be the number of these subsets. For $\idxREAC=1,\dots,n_\idxREAC$, let $\calD^{\text{disp},\idxREAC}\subseteq \calD^{\text{disp}}$ with $\calD^{\text{disp},\idxREAC} \cap \calD^{\text{disp},\idxREAC'} = \emptyset$ for $\idxREAC\neq\idxREAC'$ be the set of degrees of freedom under displacement control for which the sum of the reaction forces $\hat R^{\idxREAC,\idxLOAD}$ at time step $t$ is known. Each reaction force $\hat R^{\idxREAC,\idxLOAD}$ must be balanced by the sum of internal nodal forces on those degrees of freedom, i.e.,
\be\label{eq:weak_reaction}
\sum_{(a,i)\in\calD^{\text{disp},\beta}}F^{a,t}_i(\bftheta\RR{,\bfH}) = \sum_{(a,i)\in\calD^{\text{disp},\beta}} \int_{\partial\Omega} {\hat t}^t_i N^a(\bfX) \dd S = \hat R^{\beta,t}
\ee
Alternatively, this can be interpreted as satisfying the weak form of linear momentum balance  for the following virtual fields $\bfv$:
\be 
\eqref{eq:weak_form}\quad \text{such that}\quad \bfv = \sum\limits_{(a,i)\in\calD^{\text{disp},\beta}} N^a(\bfX){\hat\bfe}^i, \quad \forall\ \beta=1,\dots,n_\beta.
\ee

The squared residuals of the equilibrium conditions \eqref{eq:weak_bulk} and \eqref{eq:weak_reaction} can now be evaluated and added over all the time steps and corresponding degrees of freedom to obtain the cost functions described in \CFREE{} and \CDISP{} of the main article. As both $C^{\text{free}}$ and $C^{\text{disp}}$ need to be minimized to find the unknown material parameters, a multi-objective cost function (reproduced from \CTOTAL{} of the main article) is adopted
\be
C(\bftheta\RR{,\bfH}) =
C^{\text{free}}(\bftheta\RR{,\bfH})
+
\lambda_r
C^{\text{disp}}(\bftheta\RR{,\bfH})
\ee
with $\lambda_r>0$ as a hyperparameter weighting the contribution of the reaction force balance term. This weighting coefficient is important to ensure that both boundary and interior force balance terms are contributing to a similar extent to the value of the objective function. We choose $\lambda_r=100$ heuristically and keep it constant throughout the numerical experiments.

\section{Optimization of the \texorpdfstring{$\ell_p$}{lp}-regularized problem} \label{sec:optimization_details_SI}
As discussed in the main article, $\ell_p$ regularization is applied to promote sparsity in the solution vector \RR{$\bftheta$}
\be\label{eq:objective_regularized_SI}
\RR{\{\bftheta^{\text{opt}},\bfH^{\text{opt}}\} = 
\arg\min_{\{\bftheta,\bfH\geq\bf{0}\}}
}
\left(
C(\bftheta\RR{,\bfH})
+ \lambda_p\|\bftheta \|_p^p
\right),
\qquad \text{where} \qquad 
\|\bftheta\|_p = \left(\sum_{i=1}^{n_f}|\theta_i|^p\right)^{1/p}.
\ee
Evaluating the objective function for a given set of \RR{material parameters} requires calculating $\bfsigma^\idxLOAD\RR{(\bfeps^\idxLOAD,\bfh^{\idxLOAD-1},\bftheta,\bfH)}$ for each time step.
To this end, we assume $\RR{\bfh^0 = \{\bm{0},0,\bm{0}\}}$ and calculate the stress update for each time step as described in \ref{sec:return_mapping}.
After each stress update, the \RR{history variables have to be stored as they are} needed for the next step.
The implicit dependence of the cost function $C$ on the material parameters and on the internal variables calls for an optimization method that requires a small amount of cost function and gradient evaluations. We choose a trust-region reflective Newton solver \cite{coleman_convergence_1994} with gradients approximated through finite differences\footnote{The Matlab\,\textsuperscript{\tiny\textregistered} built-in optimizer \textit{lsqnonlin} is applied.}. \RR{As an alternative choice, the Nelder-Mead simplex method \cite{nelder_simplex_1965} may be applied to solve the problem. This method has shown promising results in minimizing the objective at hand, however, due to a better performance in terms of computational efficiency the trust-region reflective Newton solver was preferred.}

As a preconditioning step, we minimize $C$ without regularization with respect to the \RR{parameters $\theta_0$, $H^{\text{iso}}_1$ and $H^{\text{kin}}_1$} while constraining the other parameters to zero.
This step can be interpreted as assuming that the yield surface is described by a perfect circle in the $\pi$ plane (von Mises model) \RR{with linear isotropic and kinematic hardening}.
While being computationally inexpensive, this optimization allows to approximately estimate the size of the hidden yield surface.

\RR{Afterwards, we minimize $C$ without regularization with respect to the parameters $\bftheta$ and $\bfH$. As the trust-region reflective algorithm is not designed to find a global minimum to an optimization problem, we introduce {$n_g=100$} random initial guesses. To this end, we use $\theta_0$ from the preconditioning step, set all other parameters to zero and randomly perturb the parameters, where the perturbation follows a Gaussian distribution\footnote{\RR{Assuming the order of magnitude of the target parameters to be approximately known, the standard deviations of the Gaussian perturbations are set to $0.1/2^i$ for $\theta_i$, $1$ for $H^{\text{iso}}_2$, $100$ for $H^{\text{iso}}_1$ and $H^{\text{kin}}_1$, $1000$ for $H^{\text{iso}}_3$ and $H^{\text{kin}}_2$. In our experience, the choice of standard deviations for the random perturbations is not crucial in finding accurate solutions to the inverse problem.}}. The best solution of the unregularized minimization problem, i.e., the solution with the lowest objective function, serves then as the initial guess for the regularized minimization problem.}

The solution to \RR{the regularized problem} \eqref{eq:objective_regularized_SI} is dependent on the choice of $\lambda_p$, which controls the degree of sparsity of the solution vector.
We  propose an algorithmic procedure for automatically selecting $\lambda_p$.
To this end, we minimize \eqref{eq:objective_regularized_SI} \RR{in parallel} for a set of different choices of $\lambda_p$, specifically, \RR{$\lambda_p \in \{2^i:i=-5,\dots,15\}$}.

\begin{figure}
\begin{center}
\includegraphics[width=\textwidth]{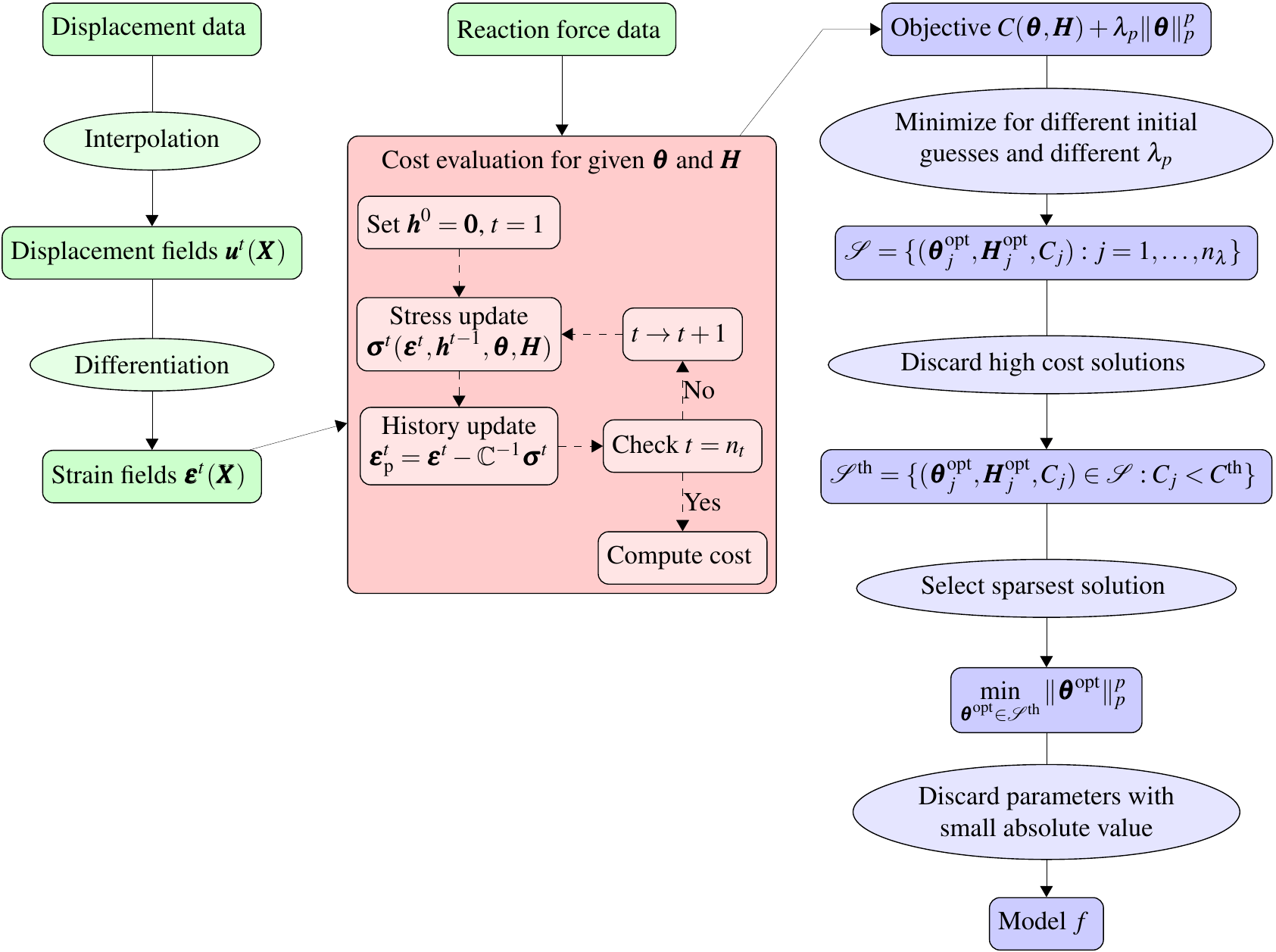}
\caption{Detailed schematics of EUCLID. Green fields correspond to data preprocessing, red fields correspond to cost function evaluation and blue fields correspond to optimization and model discovery.}
\label{fig:flowchart}
\end{center}
\end{figure}

We hence obtain one solution for each choice of $\lambda_p$, which we store in the set of possible solutions
\be
\calS = \{(\bftheta^{\text{opt}}_j\RR{,\bfH^{\text{opt}}_j},C_j) : j=1,\dots,n_\lambda\},
\ee
where $n_\lambda$ is the number of different choices of $\lambda_p$.
The solution corresponding to the lowest cost in the solution set $\calS$ is expected to feature a dense solution vector and to provide the highest accuracy.
To find a parsimonious material model which still provides a reasonable fitting accuracy, we first discard all solutions whose cost is higher than a threshold value $C^{\text{th}}$
\be
\calS^{\text{th}} = \{(\bftheta^{\text{opt}}_j\RR{,\bfH^{\text{opt}}_j},C_j)\in\calS : C_j < C^{\text{th}}\}.
\ee
From the remaining solutions $\calS^{\text{th}}$, which are expected to provide an acceptable accuracy if $C^{\text{th}}$ is sufficiently low, the sparsest solution is selected by choosing the solution with the smallest $\|\bftheta^{\text{opt}}_j\|_p^p$.
As the $\ell_{p}$ regularization shrinks certain parameters but does not exactly set them to zero, a final thresholding is applied, where all parameters in the selected solution vector with absolute value below a threshold $\theta^{\text{th}}$ are set to zero.
Note that the parameters $C^{\text{th}}$ and $\theta^{\text{th}}$ are the only hyperparameters in the proposed selection rule, resulting in low effort for parameter tuning.
We here choose $C^{\text{th}}$ to be slightly greater than the lowest cost $C^{\text{min}}$ in the solution set $\calS$, i.e., $C^{\text{th}} = 1.01 \ C^{\text{min}}$, \RR{and $\theta^{\text{th}}$ to be much smaller than the first component of the selected solution vector, i.e., $\theta^{\text{th}}=0.005 \ \theta_0$}. \figurename~\ref{fig:flowchart} schematically summarizes the optimization process and \tablename~\ref{tab:settings_SI} lists the corresponding parameters.

\begin{figure}
\begin{center}
\includegraphics[width=0.6\textwidth]{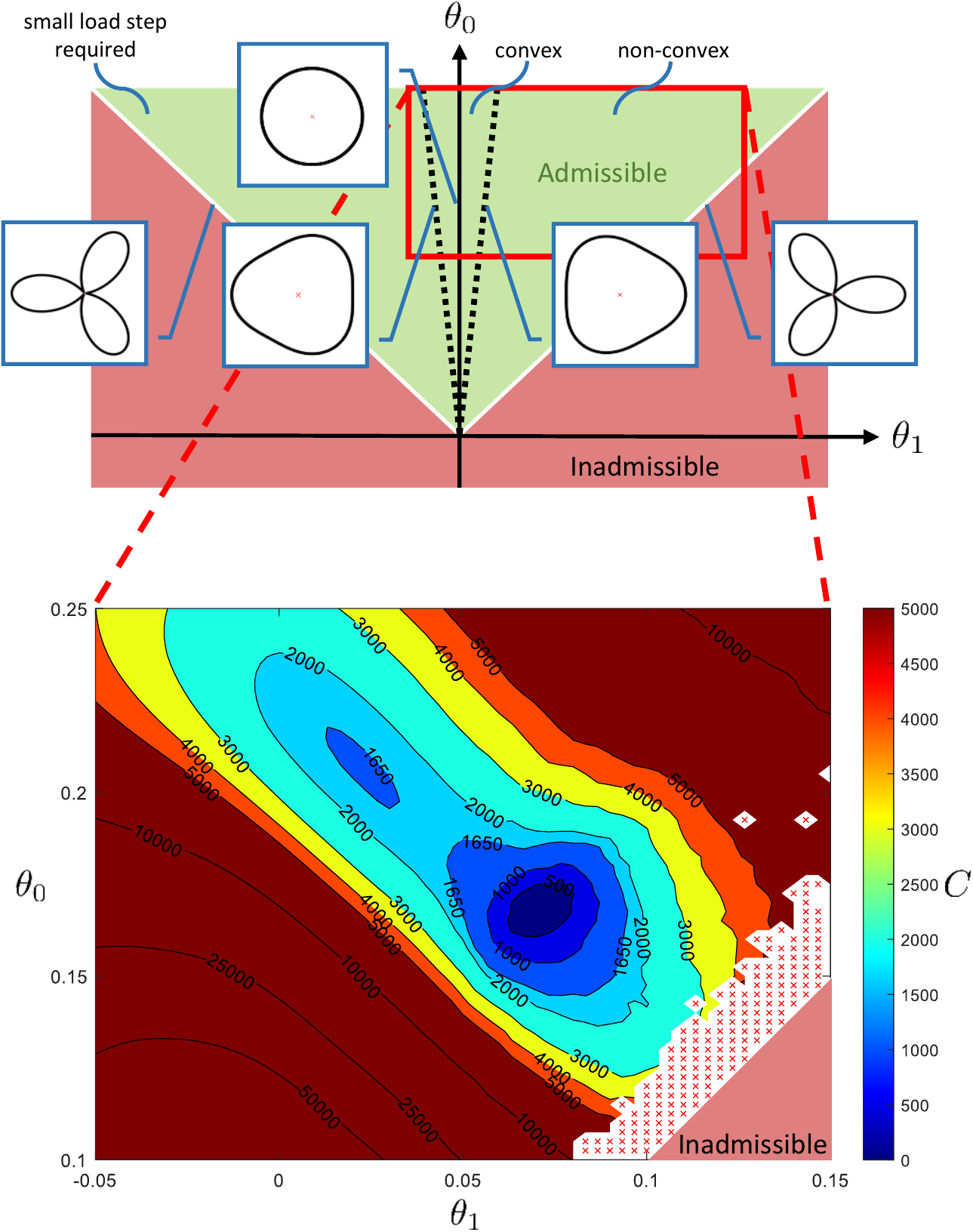}
\caption{Top: Illustration of the physical requirements on the parameters $\bftheta$ and exemplary shapes of yield surfaces. Only the projection of $\bftheta$ on the $\theta_1$-$\theta_0$ plane \RR{under the assumption of perfect plasticity ($\bfH = \bm{0}$)} is shown. The yield surface is physically admissible if $\theta_0 > |\theta_1|$ (necessary condition; green area) and inadmissible otherwise (red area). The dashed lines at $\theta_0 = 10|\theta_1|$ separate convex from non-convex yield surfaces. Bottom: Cost function $C(\bftheta)$ evaluations for different values of $\theta_0$ and $\theta_1$ (assuming the rest of the features to be zero) for the \RR{perfectly plastic version of the non-convex yield surface, i.e., model \ModelNonConvex{} with $\bfH = \bm{0}$}. There is a local minimum distinct from the global minimum. The cost function evaluation fails near the physically inadmissible region (red crosses) due to convergence issues of the return mapping algorithm. Material parameters $\bftheta$ are in $\text{kN}/\text{mm}^2$ and the cost is in $\text{kN}$.}
\label{fig:theta_space}
\end{center}
\end{figure}

Each optimization process requires to repeatedly evaluate the cost function, which in turn requires to repeatedly apply the stress-update procedure. Depending on the load step size and the parameters for which the cost function is evaluated, the stress update procedure may fail to converge. In such scenario the corresponding cost function is set to infinity. This can hinder the trust-region reflective algorithm to converge to a local minimum, in which case the corresponding cost is set to infinity. This issue is observed to be particularly dominant if the parameters for which the cost function is evaluated get closer to violating the physical constraint in \CONSTRAINTS{} of the main article (see \figurename~\ref{fig:theta_space}). \RR{In our numerical examples this phenomenon only occurred during the optimization of the unregularized problem for some of the random initial guesses. The convergence issues were not present in the regularized problem, for which we used the converged solution from the unregularized problem as initial guess.} Note that when running a forward problem, the convergence issues can be fixed by decreasing the load step size. In the inverse problem, however, the load step size is predefined by the given data set. Synthetically generating data points by linear interpolation between subsequent points in time may be a solution to overcome this problem, but was not tried in the present work.

\clearpage

\clearpage

\clearpage

\clearpage

\clearpage

\begin{figure}
\begin{center}
\includegraphics[width=0.8\textwidth]{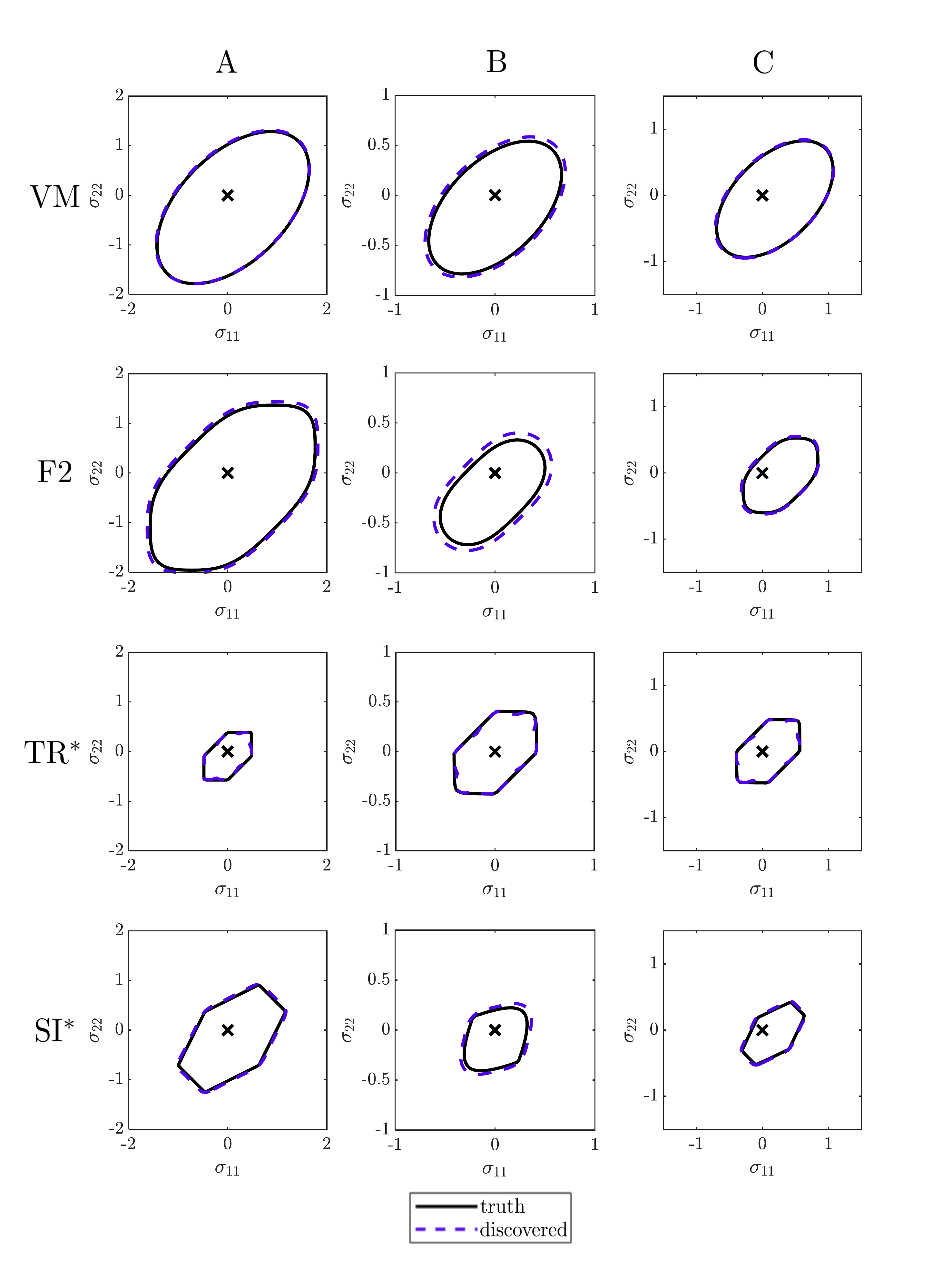}
\caption{\RR{Yield surface plots of the (true) hidden and discovered plasticity models \ModelVonMisesTrue, \ModelSparseTwo, \ModelTresca, \ModelSchmidtIshlinsky{} at the end of three different deformation paths (A, B, C) corresponding to characteristic points in the specimen domain (see \figurename~4 in the main article). A noise level of $\sigma=10^{-4}\text{mm}$ was considered and the yield surfaces are plotted at $\sigma_{12}=0$. $\sigma_{ij}$ are in $\text{kN}/\text{mm}^2$.}}
\label{fig:yield_surfaces_kin_hardening_1234}
\end{center}
\end{figure}

\clearpage

\begin{figure}
\begin{center}
\includegraphics[width=0.8\textwidth]{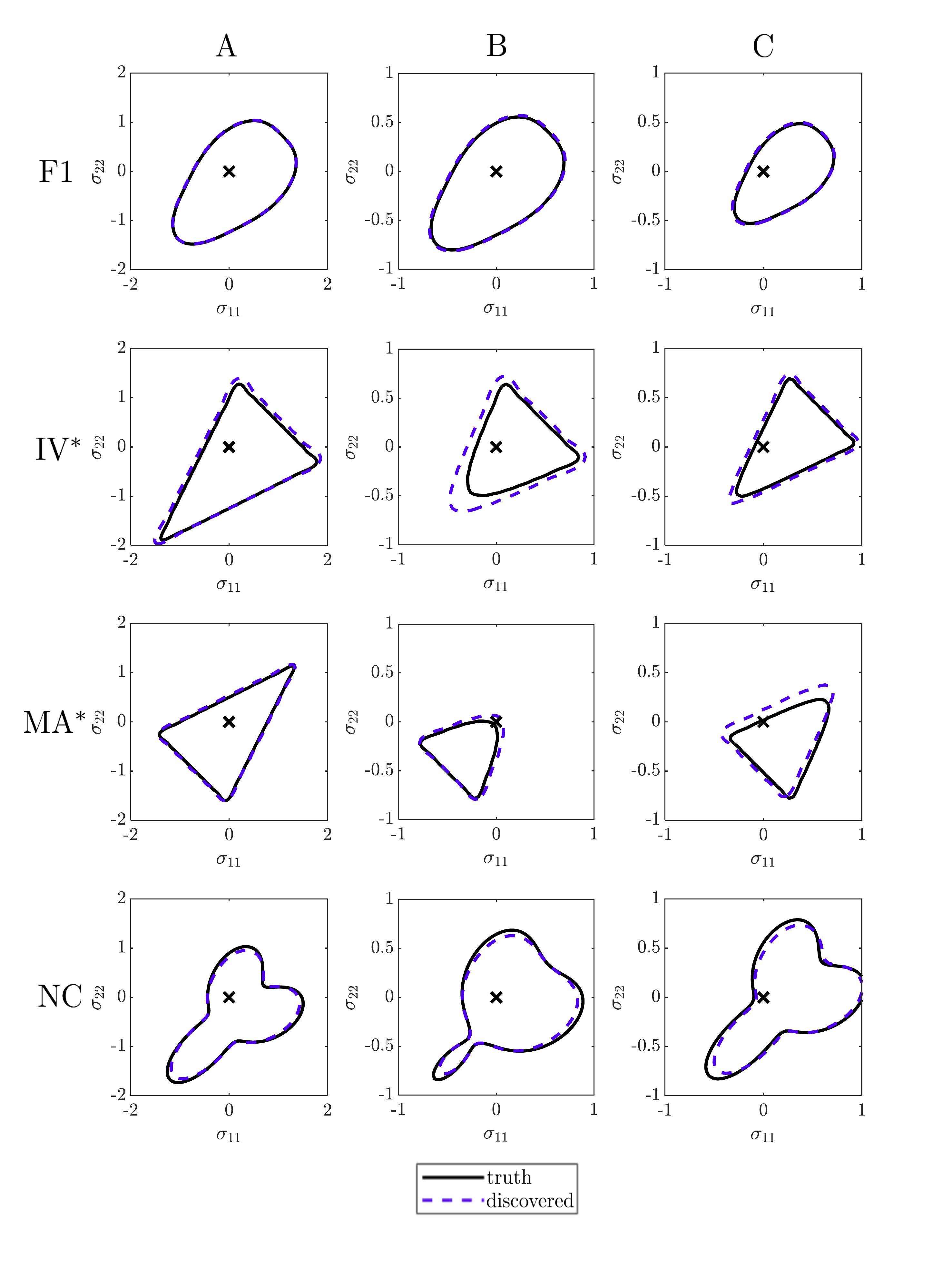}
\caption{\RR{Yield surface plots of the (true) hidden and discovered plasticity models \ModelSparseOne, \ModelIvlev, \ModelMariotte, \ModelNonConvex{} at the end of three different deformation paths (A, B, C) corresponding to characteristic points in the specimen domain (see \figurename~4 in the main article). A noise level of $\sigma=10^{-4}\text{mm}$ was considered and the yield surfaces are plotted at $\sigma_{12}=0$ for load paths A and C and $\sigma_{12}=0.25$ for load path B. $\sigma_{ij}$ are in $\text{kN}/\text{mm}^2$.}}
\label{fig:yield_surfaces_kin_hardening_5678}
\end{center}
\end{figure}

\clearpage

\begin{table}
\centering
\caption{\RR{Non-smooth yield functions and their fully expanded smooth approximations. For better clarity we show the initial yield functions, i.e., $\gamma=0$.} Material parameters are in $\text{kN}/\text{mm}^2$.}
\label{tab:material_models_approximations_SI}
\begin{tabular}{lll}
\hline
\multicolumn{2}{c}{Benchmarks} & \multicolumn{1}{l}{Yield Function $f$} \\%
\hline\\[-10pt]
\ModelTrescaTrue & Truth & $\max(|\sigma_1-\sigma_2|,|\sigma_2-\sigma_3|,|\sigma_3-\sigma_1|) - 0.2400$\\%
\ModelTresca & Truth$^*$ & $\sqrt{3/2} r - \big( 0.2181 + 0.0127 \cos(6\alpha) + 0.0035 \cos(12\alpha) + 0.0016 \cos(18\alpha)$\\%
 &  & \hphantom{$\sqrt{3/2} r - \big($} $+$ $0.0009 \cos(24\alpha) + 0.0006 \cos(30\alpha) + 0.0004 \cos(36\alpha) + 0.0003 \cos(42\alpha)$\\%
 &  & \hphantom{$\sqrt{3/2} r - \big($} $+$ $0.0002 \cos(48\alpha) + 0.0002 \cos(54\alpha) + 0.0001 \cos(60\alpha) \big)$\\%
\hline\\[-10pt]
\ModelSchmidtIshlinskyTrue & Truth & $\max(|\sigma_1-(\sigma_2+\sigma_3)/2|,|\sigma_2-(\sigma_3+\sigma_1)/2|,|\sigma_3-(\sigma_1+\sigma_2)/2|) - \cos(\pi / 6) 0.2400$\\%
\ModelSchmidtIshlinsky & Truth$^*$ & $\sqrt{3/2} r - \big( 0.2193 - 0.0128 \cos(6\alpha) + 0.0035 \cos(12\alpha) - 0.0016 \cos(18\alpha)$\\%
 &  & \hphantom{$\sqrt{3/2} r - \big($} $+$ $0.0009 \cos(24\alpha) - 0.0006 \cos(30\alpha) + 0.0004 \cos(36\alpha) - 0.0003 \cos(42\alpha)$\\%
 &  & \hphantom{$\sqrt{3/2} r - \big($} $+$ $0.0002 \cos(48\alpha) - 0.0002 \cos(54\alpha) + 0.0001 \cos(60\alpha) \big)$\\%
\hline\\[-10pt]
\ModelIvlevTrue & Truth & $\max((\sigma_2+\sigma_3)-2\sigma_1,(\sigma_3+\sigma_1)-2\sigma_2,(\sigma_1+\sigma_2)-2\sigma_3) - 0.2400$\\%
\ModelIvlev & Truth$^*$ & $\sqrt{3/2} r - \big( 0.1509 + 0.0415 \cos(3\alpha) + 0.0157 \cos(6\alpha) + 0.0081 \cos(9\alpha)$\\%
 &  & \hphantom{$\sqrt{3/2} r - \big($} $+$ $0.0049 \cos(12\alpha) + 0.0032 \cos(15\alpha) + 0.0023 \cos(18\alpha) + 0.0017 \cos(21\alpha)$\\%
 &  & \hphantom{$\sqrt{3/2} r - \big($} $+$ $0.0013 \cos(24\alpha) + 0.0011 \cos(27\alpha) + 0.0009 \cos(30\alpha) \big)$\\%
\hline\\[-10pt]
\ModelMariotteTrue & Truth & $\max(\sigma_1-(\sigma_2+\sigma_3)/2,\sigma_2-(\sigma_3+\sigma_1)/2,\sigma_3-(\sigma_1+\sigma_2)/2) - \cos(\pi / 3) 0.2400$\\%
\ModelMariotte & Truth$^*$ & $\sqrt{3/2} r - \big( 0.1563 - 0.0430 \cos(3\alpha) + 0.0163 \cos(6\alpha) - 0.0084 \cos(9\alpha)$\\%
 &  & \hphantom{$\sqrt{3/2} r - \big($} $+$ $0.0051 \cos(12\alpha) - 0.0034 \cos(15\alpha) + 0.0024 \cos(18\alpha) - 0.0018 \cos(21\alpha)$\\%
 &  & \hphantom{$\sqrt{3/2} r - \big($} $+$ $0.0014 \cos(24\alpha) - 0.0011 \cos(27\alpha) + 0.0009 \cos(30\alpha) \big)$\\%
\hline\\[-10pt]
\end{tabular}%
\end{table}


\begin{table}
	\caption{Default parameters and hyperparameters for FEM data generation and EUCLID.}
	\label{tab:settings_SI}
	\centering
	\begin{tabular}{lccc}
		\hline
		\multicolumn{4}{c}{FEM Data Generation}\\\hline
		Parameter & Notation & Value & Unit \\ \hline
		Number of nodes in mesh & $n_n$ & $2,179$ & - \\
		Number of reaction force constraints & $n_\idxREAC$ & $2$ & - \\
		Number of time steps & $n_t$ & \RR{2250} & - \\
		Loading parameter (tension phase) & $\delta$ & $[0,\RR{0.5}]$ & mm \\
		Loading parameter (compression phase) & $\delta$ & $[\RR{0.5},-\RR{0.5}]$ & mm \\
		Young's modulus & $E$ & $210$ & kN/mm$^2$ \\
		Poisson's ratio & $\nu$ & $0.3$ & - \\
		Displacement noise standard deviation & $\sigma$ & $\left\{0,10^{-4},3 \cdot 10^{-4},5 \cdot 10^{-4}\right\}$ & mm\\
		Denoising moving-window length & - & $50$ & -\\
		\hline
		\\
		\\\hline
		\multicolumn{4}{c}{EUCLID}\\\hline
		Parameter & Notation & Value & Unit \\ \hline
		Number of features & $n_f+1$ & \RR{7} & - \\
		Coefficient for reaction force balance & $\lambda_r$ & 100 & - \\
		Exponent for $\ell_p$ regularization & $p$ & $1/4$ & - \\
		Coefficient for $\ell_p$ regularization & $\lambda_p$ & $\{2^i:i=\RR{-5,\dots,15}\}$ & $\text{mm}^{2p}/\text{kN}^{p-1}$ \\
		Number of choices of $\lambda_p$ & $n_\lambda$ & \RR{21} & -\\
		Number of random initial guesses & $n_g$ & $100$ & -\\
		Threshold for $C$ & $C^{\text{th}}$ & $0.01 \ C^{\text{min}}$ & kN\\
		Threshold for $\bftheta$ & $\theta^{\text{th}}$ & $0.005 \ \theta_0$ & kN/mm$^2$\\
		
		\hline
	\end{tabular}
\end{table}

\end{document}